\documentclass[a4paper,fleqn]{cas-sc}

\usepackage[authoryear,longnamesfirst]{natbib}
\usepackage{graphicx}
\usepackage{amsmath,amssymb}
\usepackage{url}
\usepackage{booktabs}
\usepackage{tikz}
\usetikzlibrary{shapes,arrows,positioning,fit,backgrounds,calc}
\usepackage{tikz}
\usepackage{amsmath}
\usepackage{paralist}
\usepackage{hyperref}
\usepackage{adjustbox}
\usepackage{filecontents}

\def\tsc#1{\csdef{#1}{\textsc{\lowercase{#1}}\xspace}}
\tsc{WGM}
\tsc{QE}

\usepackage{soul}
\newcommand{\changed}[1]{{#1}}
\newcommand{\changedr}[2]{{#2}}
\newcommand{\changedold}[1]{{#1}}

\begin{document}
\let\WriteBookmarks\relax
\def\floatpagepagefraction{1}
\def\textpagefraction{.001}

\shorttitle{SemVul: Semantic-Enhanced Graph Neural Networks for Code Property Graph-based Vulnerability Detection}    

\shortauthors{Affan et al.}  

\title [mode = title]{SemVul: Semantic-Enhanced Graph Neural Networks for Code Property Graph-based Vulnerability Detection}

\author[1]{Younas Affan}[orcid=0009-0001-9628-7602]
\ead{younas.affan@unica.it}
\author[1]{Leonardo Regano}[orcid=0000-0002-9259-5157]
\ead{leonardo.regano@unica.it}
\cormark[1]
\author[1,2]{Giorgio Giacinto}[orcid=0000-0002-5759-3017]
\ead{giacinto@unica.it}
\affiliation[1]
{
organization={Dipartimento di Ingegneria Elettrica e Elettronica, Universit\`a di Cagliari},
addressline={Via Marengo 3},
postcode={09123},
city={Cagliari},
country={Italy}
}
\affiliation[2]{
organization={Interuniversity National Consortium for Informatics, CINI},
addressline={Via Ariosto, 25}, 
city={Rome},
postcode={00185}, 
country={Italy}
}
\cortext[1]{Corresponding author}

\begin{abstract}
Vulnerabilities in source code are often the root cause of cyberattacks worldwide, as attackers exploit weaknesses in software to gain unauthorized access, steal data, or disrupt services. In this study, we evaluated existing research approaches and propose SemVul, a \changedr{new method}{vulnerability detection pipeline} that demonstrates better generalization and higher accuracy in learning vulnerable code patterns. We propose a Code Property Graph-based vulnerability-detection approach combined with semantic-level enhancement, enabling the model to capture both the program's structural flow and the semantic meaning of the code. Our approach integrates both node-level and edge-level semantic embeddings using pre-trained code embedding techniques. We systematically evaluate multiple GNN architectures on publicly available benchmark datasets. \changedr{We present a new source code vulnerability detection pipeline that}{SemVul} is generic with respect to the programming language and supports multiple architectures. By integrating structural and semantic information, the proposed approach improves vulnerability detection performance. Our results show that SemVul outperforms existing approaches and provides better generalization.
\end{abstract}

\begin{highlights}
\item Semvul is a language-agnostic, reproducible pipeline for code vulnerability detection.
\item Code vulnerability detection methods often struggle to generalise across benchmarks.
\item SemVul unifies the Code Property Graph structure with pre-trained code semantics.
\item Combining structural and semantic code representations improves generalisation.
\item SemVul matches or surpasses performances of prior methods across benchmarks.
\end{highlights}


\maketitle

\section{Introduction}
\label{sec:introduction}

Software vulnerabilities remain a critical threat to modern computing systems. In 2024 alone, over 25,000 new vulnerabilities were reported in the CVE database~\cite{lekssays2025llmxcpg}, while the open-source ecosystem continues its explosive growth with more than 60 million new repositories created on GitHub in 2020~\cite{song2022hgvul}. This scale amplifies the attack surface and enables supply-chain attacks, as exemplified by high-profile incidents like SolarWinds. Vulnerabilities introduced at the source code level propagate across dependent projects, leading to severe security breaches. Preventing such exploits requires accurate and scalable detection early in the development lifecycle.

Existing approaches, such as static analysis, dynamic analysis, and symbolic execution, provide formal guarantees for certain vulnerability classes but suffer from well-known limitations. Static analyzers produce high false-positive rates and rely on expert-written rules, while dynamic approaches impose significant computational overhead, unsuitable for continuous integration pipelines~\cite{marjanov2022mlreview}. This has motivated growing interest in data-driven techniques that \emph{learn} vulnerability patterns directly from source code.

\subsection{Achievements and Limitations of Prior Work}

We categorize existing learning-based approaches into three paradigms, each with distinct strengths and weaknesses.

\textbf{Token-based sequence models} are used by methods such as VulDeePecker~\cite{li2018vuldeepecker}, SySeVR~\cite{li2021sysevr}, and LineVul~\cite{fu2022linevul}, which treat code as token sequences processed by neural networks. LineVul leverages CodeBERT pretraining~\cite{feng2020codebert} to achieve strong results on curated benchmarks, capturing contextual token patterns that correlate with vulnerabilities. However, these models fundamentally ignore program structure. Recent evaluations reveal that accuracy drops by more than 50\% when synthetic and duplicated samples are removed~\cite{chakraborty2022arewethere}. The Real-Vul benchmark shows precision losses up to 95\% when evaluating on full projects rather than isolated functions~\cite{chakraborty2024realvul}. This suggests token-based models learn superficial lexical patterns rather than meaningful vulnerability semantics.

A second approach is based on \textbf{graph-based structural models}. Devign~\cite{zhou2019devign} pioneered using Code Property Graphs (CPGs) with gated graph neural networks, demonstrating significant gains over sequence models. Follow-up work extends this: HGVul~\cite{song2022hgvul} uses heterogeneous graphs with attention mechanisms; DeepWukong~\cite{cheng2021deepwukong} applies program slicing; IVDetect~\cite{li2021vulnerability} combines control-flow, data-flow, and call graphs. These methods consistently outperform token-based baselines by capturing structural program dependencies. However, graph-based models often underutilize semantic code content. By relying on one-hot node type encodings or simple feature engineering, they miss the rich contextual meaning encoded in identifiers, literals, and API names. Additionally, GNNs face challenges with over-smoothing~\cite{li2018oversmoothing} and over-squashing, limiting their ability to propagate vulnerability-relevant signals across complex graphs.

Finally, another approach leverages \textbf{Large Language Models}. Indeed, LLMs like GPT-5 and CodeLlama offer impressive code understanding capabilities, potentially leveraging vast pretraining knowledge to identify vulnerability patterns. However, comprehensive evaluations caution against relying too heavily on LLMs for security tasks. The SecLLMHolmes framework~\cite{ullah2024secllmholmes} found non-deterministic responses and high false-positive rates across eight LLMs. Simple changes such as renaming variables caused PaLM2 and GPT-4 to misclassify 26\% and 17\% of examples, respectively. Furthermore, the PrimeVul benchmark~\cite{ding2024primevul} demonstrated that GPT-4's vulnerability predictions perform \emph{worse than random guessing} on paired evaluation, with balanced accuracy of only 46.3\% below the 50\% random baseline. These results reveal fundamental brittleness that makes LLMs unreliable for security-critical applications.

\subsection{Our Approach: Semantic-Enhanced Graph Learning}

Motivated by these observations, we propose \textbf{SemVul}, a semantic-enhanced \changedr{graph neural network}{vulnerability detection pipeline} that addresses the limitations of both token-based and purely structural approaches. Our key insight is that effective vulnerability detection requires \emph{both} structural understanding of program flow \emph{and} semantic understanding of code content.

\begin{figure}
\centering
\resizebox{\textwidth}{!}{\begin{tikzpicture}[
    node distance=0.5cm and 0.7cm,
    box/.style={rectangle, draw, rounded corners, minimum height=0.9cm, minimum width=1.8cm, align=center},
    data/.style={box, fill=blue!10},
    process/.style={box, fill=green!10},
    model/.style={box, fill=orange!10},
    output/.style={box, fill=red!10},
    cpgbox/.style={box, fill=cyan!10, minimum width=2.8cm, minimum height=2cm},
    arrow/.style={->, thick, >=stealth},
    label/.style={font=\itshape, text=gray}
]

\node[data, minimum width=2cm] (source) {Source Code\\(C/C++)};

\node[process, right=0.5cm of source] (joern) {Joern\\Parser};

\node[cpgbox, right=0.5cm of joern] (cpg) {};
\node[font=\bfseries, text=cyan!50!black] at ($(cpg.north)-(0,-0.2)$) {CPG (GraphSON)};
\node[data, minimum width=1.2cm, minimum height=0.5cm] at ($(cpg.center)+(-0,0.4)$) (nodes) {Nodes};
\node[data, minimum width=1.2cm, minimum height=0.5cm] at ($(cpg.center)+(0,-0.3)$) (edges) {Edges};

\coordinate (cpgsplit) at ($(cpg.east)+(0.3,0)$);

\node[process, minimum width=1.6cm] (exttype) at ($(cpgsplit)+(1.2,0.8)$) {Extract\\Node Type};
\node[process, minimum width=1.6cm] (extcode) at ($(cpgsplit)+(1.2,-0.8)$) {Extract\\CODE prop};

\node[process, fill=blue!15, minimum width=1.4cm] (onehot) at ($(cpgsplit)+(3.2,0.8)$) {One-Hot\\Encode};
\node[process, fill=purple!15, minimum width=1.4cm] (embed) at ($(cpgsplit)+(3.2,-0.8)$) {Semantic\\Embed};

\node[process, fill=orange!15, minimum width=1.6cm] (concat) at ($(cpgsplit)+(5.6,0)$) {Concatenate\\$[T|D]$};
\coordinate (concat_top) at ($(concat.west)+(0,0.22)$);
\coordinate (concat_bot) at ($(concat.west)+(0,-0.22)$);

\node[data, fill=yellow!15, minimum width=1.6cm] (pyg) at ($(cpgsplit)+(7.8,0)$) {PyG\\Data};

\node[model, minimum width=1.6cm] (gnn) at ($(cpgsplit)+(9.8,0)$) {GNN\\Layers};

\node[model, minimum width=1.4cm] (pool) at ($(cpgsplit)+(11.6,0)$) {Pooling};

\node[output, minimum width=1.4cm] (class) at ($(cpgsplit)+(13.4,0)$) {Vuln?\\0/1};

\draw[arrow] (source.east) -- (joern.west);
\draw[arrow] (joern.east) -- (cpg.west);
\draw[arrow] (cpg.east) -- (cpgsplit);
\draw[arrow] (cpgsplit) |- (exttype.west);
\draw[arrow] (cpgsplit) |- (extcode.west);
\draw[arrow] (exttype.east) -- (onehot.west);
\draw[arrow] (extcode.east) -- (embed.west);
\draw[arrow] (onehot.east) -- (concat_top);
\draw[arrow] (embed.east) -- (concat_bot);
\draw[arrow] (concat.east) -- (pyg.west);
\draw[arrow] (pyg.east) -- (gnn.west);
\draw[arrow] (gnn.east) -- (pool.west);
\draw[arrow] (pool.east) -- (class.west);

\end{tikzpicture}}
\caption{SemVul system architecture. Source code is parsed by Joern into a Code Property Graph (CPG) stored as GraphSON. Each CPG node contains properties including \texttt{CODE} (source code text) and \texttt{label} (node type). We extract node types for one-hot encoding ($T$ dimensions) and \texttt{CODE} properties for semantic embedding ($D$ dimensions). These are concatenated to form node features, combined with graph structure into PyG Data, processed by GNN layers, pooled to graph-level, and classified as vulnerable or safe.}
\label{fig:system_architecture}
\end{figure}
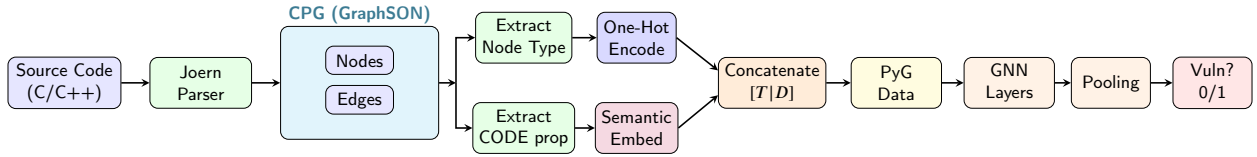

As illustrated in Figure~\ref{fig:system_architecture}, SemVul operates as follows:

\begin{enumerate}
    \item \textbf{Code Property Graph Extraction:} We use Joern~\cite{yamaguchi2014cpg} to parse source functions into CPGs that unify AST, CFG, and DFG representations into a single heterogeneous graph. Each CPG node corresponds to a source code element (statement, expression, or declaration), with source code text extracted from the CPG's \texttt{CODE} property.

    \item \textbf{Semantic Embedding Integration:} Each CPG node contains source code text (extracted from the graph's \texttt{CODE} property), which we transform into dense vector representations using pre-trained code models (CodeBERT, GraphCodeBERT, UniXcoder) or dataset-specific Word2Vec. This captures the semantic meaning of identifiers, API calls, and literals information invisible to purely structural approaches. The CPG edges then serve to \emph{correlate} these source code embeddings: data-flow edges connect statements where a variable's definition reaches its use, control-flow edges link statements in execution order, and syntactic edges preserve hierarchical code structure.

    \item \textbf{Edge-Type-Aware GNN Processing:} We employ GNN architectures that explicitly model different edge types (syntax, control-flow, data-flow), allowing the model to learn which relationships between source code lines are most relevant for detecting specific vulnerability patterns. For instance, data-flow edges help trace tainted values from input to dangerous operations, while control-flow edges reveal missing safety checks.

    \item \textbf{Multi-Statistic Graph Pooling:} We aggregate node representations into a graph-level vector using pooling strategies that capture distributional properties (mean, max, standard deviation), providing richer signals than single-statistic approaches.
\end{enumerate}

\begin{figure}
\centering
\begin{tikzpicture}[
    node distance=0.5cm and 0.4cm,
    codenode/.style={rectangle, draw, rounded corners=2pt, minimum height=0.55cm, minimum width=1.2cm, align=center, font=\scriptsize\ttfamily},
    astnode/.style={codenode, fill=blue!15},
    cfgnode/.style={codenode, fill=green!15},
    callnode/.style={codenode, fill=red!20, line width=0.8pt},
    astedge/.style={->, blue!60, line width=0.6pt},
    cfgedge/.style={->, green!50!black, line width=0.6pt, dashed},
    dfgedge/.style={->, red!60, line width=0.6pt, dotted},
    codebox/.style={rectangle, draw, fill=gray!5, align=left, font=\scriptsize\ttfamily, inner sep=6pt}
]

\node[codebox] (code) {
\textbf{void} vulnerable(\textbf{char} *input) \{\\
\quad \textbf{char} buffer[10];\\
\quad strcpy(buffer, input); \textcolor{red}{// CWE-120}\\
\}
};

\node[below=0cm of code] (arrow) {$\Downarrow$ \textit{Joern CPG}};

\node[astnode, below=0cm of arrow] (method) {METHOD\\vulnerable};
\node[astnode, below left=0.5cm and 0.6cm of method] (param) {PARAM\\input};
\node[astnode, below right=0.5cm and 0.6cm of method] (local) {LOCAL\\buffer[10]};
\node[callnode, below=1.4cm of method] (strcpy) {CALL\\strcpy};
\node[astnode, below left=0.5cm and 0.2cm of strcpy] (arg1) {ARG\\buffer};
\node[astnode, below right=0.5cm and 0.2cm of strcpy] (arg2) {ARG\\input};
\node[cfgnode, below=0.9cm of strcpy] (ret) {RETURN};

\draw[astedge] (method) -- (param);
\draw[astedge] (method) -- (local);
\draw[astedge] (method) -- (strcpy);
\draw[astedge] (strcpy) -- (arg1);
\draw[astedge] (strcpy) -- (arg2);

\draw[cfgedge] (method) to[bend left=20] (local);
\draw[cfgedge] (local) -- (strcpy);
\draw[cfgedge] (strcpy) -- (ret);

\draw[dfgedge, line width=0.8pt] (param) to[bend right=30] (arg2);
\draw[dfgedge] (local) to[bend left=20] (arg1);

\node[below=0.2cm of ret, align=center] (legend) {
\tikz[baseline=-0.5ex]\draw[blue!60, line width=0.6pt, ->] (0,0) -- (0.4,0); AST \quad
\tikz[baseline=-0.5ex]\draw[green!50!black, line width=0.6pt, dashed, ->] (0,0) -- (0.4,0); CFG \quad
\tikz[baseline=-0.5ex]\draw[red!60, line width=0.6pt, dotted, ->] (0,0) -- (0.4,0); DFG
};

\end{tikzpicture}
\caption{Code-to-graph transformation for buffer overflow (CWE-120). Untrusted \texttt{input} flows via DFG edges to \texttt{strcpy}, which writes to a fixed buffer.}
\label{fig:code_to_graph}
\end{figure}
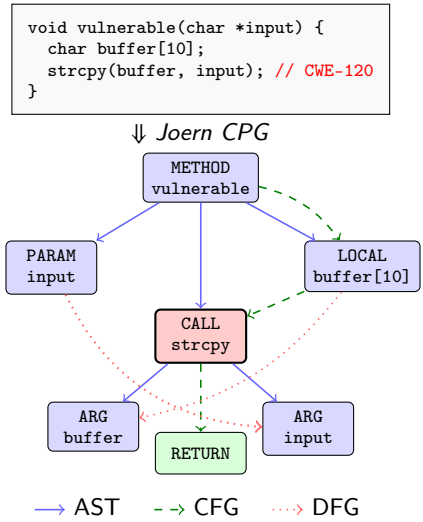

\newpage
\changedold{Throughout the rest of the paper we use the function of
Figure~\ref{fig:code_to_graph} as a running example. The function copies an
attacker-controlled parameter \texttt{input} into a fixed ten-byte
\texttt{buffer} via \texttt{strcpy}, a classic buffer overflow (CWE-120).
Joern turns it into a CPG with seven nodes (\texttt{METHOD}, \texttt{PARAM
input}, \texttt{LOCAL buffer}, \texttt{CALL strcpy}, its two \texttt{ARG}
nodes, and \texttt{RETURN}) connected by three kinds of typed edges: AST edges
give the syntactic backbone, CFG edges order the declaration, the call, and
the return, and DFG edges carry the vulnerability-relevant fact that
\texttt{input} flows unmodified into the second argument of \texttt{strcpy}.}

Our approach addresses the core limitations identified in prior work: by combining CPG structure with code embeddings, we capture both ``how data flows'' (structure) and ``what the code means'' (semantics). A dangerous function call like \texttt{strcpy} is recognized both by its position in the data-flow graph and by its semantic embedding.

Furthermore, unlike token-based or LLM approaches that are sensitive to variable renaming, our structural representation remains stable under cosmetic code changes. The CPG topology is preserved even when identifiers are renamed.

Finally, different vulnerability types manifest in different program aspects. Buffer overflows are visible in data flow; null pointer dereferences appear in control flow. \changedr{Our multi-relational GNN architectures (RGAT, HGT, MultiView) can learn}{We employ multi-relational GNN architectures (RGAT, HGT, MultiView), able to learn which edge types are most discriminative for a given vulnerability pattern.}

\subsection{Contributions}

This paper makes the following contributions:

\begin{enumerate}
    \item \textbf{Semantic-Enhanced CPG Learning:} We propose integrating pre-trained code embeddings with Code Property Graphs, enabling GNNs to jointly model structural program flow and semantic code meaning.

    \item \textbf{Comprehensive Architecture Comparison:} We systematically evaluate multiple GNN architectures (GGNN, RGAT, HGT, MultiView, Call-Aware) and embedding strategies (Word2Vec, CodeBERT, UniXcoder) to understand their contributions to detection performance.

    \item \textbf{Reproducible End-to-End Pipeline:} We provide a complete, reproducible pipeline from raw source code to trained models, with explicit handling of data quality issues including deduplication and cross-split leakage prevention.

    \item \textbf{State-of-the-Art Results:} SemVul achieves competitive or superior performance on multiple benchmarks, outperforming prior methods.

    \item \textbf{Language-Agnostic Architecture:} While our experiments focus on C/C++ code, the pipeline architecture is language-agnostic. Joern supports parsing of multiple programming languages (Java, JavaScript, Python, Go, among others), and our graph construction and GNN components require no language-specific modifications. Extending to other languages requires only retraining embeddings and regenerating CPGs the model architecture remains unchanged.
\end{enumerate}

\subsection{Research Questions}

Our study addresses the following research questions:

\begin{itemize}

    \item \textbf{RQ1 (Semantic Enhancement):} \emph{To what extent do pre-trained code embeddings (CodeBERT, GraphCodeBERT, UniXcoder) improve vulnerability detection effectiveness measured by F1 score (balance between correctly identifying vulnerabilities and avoiding false alarms) and AUC (ability to rank vulnerable code higher than safe code) over baseline representations (one-hot node types, dataset-specific Word2Vec)?}

    We hypothesize that pre-trained embeddings capture richer semantic information than dataset-specific or structural-only representations. We measure the absolute improvement in F1 and AUC when replacing one-hot encodings with progressively richer embeddings.

    \item \textbf{RQ2 (GNN Architecture):} \emph{Among GGNN, RGAT, HGT, MultiView, and Call-Aware architectures, which achieves the best vulnerability detection performance (highest F1 score) on each benchmark, and do edge-type-aware models consistently outperform edge-agnostic baselines?}

    We systematically compare architectures across four datasets (Devign~\cite{zhou2019devign}, Reveal~\cite{chakraborty2022arewethere}, Draper~\cite{russell2018automated}, PrimeVul~\cite{ding2024primevul}) to determine whether explicit edge-type modeling provides consistent benefits. We also examine whether architecture choice interacts with dataset characteristics (e.g., class imbalance, graph size).

    \item \textbf{RQ3 (Cross-Dataset Transfer):} \emph{What is the degradation in vulnerability detection capability (measured by F1 drop) when models trained on one dataset are evaluated on another, and which architectural choices minimize this degradation?}

    We evaluate cross-dataset transfer by training on each dataset and testing on the others. This quantifies the generalization gap caused by schema heterogeneity (15-16 node types, 18-19 edge types) and distribution shift, providing practical guidance for deployment scenarios.

    \item \textbf{RQ4 (Multi-Dataset Generalization):} \emph{Does SemVul maintain consistent vulnerability detection performance across datasets with diverse characteristics (balanced vs.\ imbalanced, noisy vs.\ clean labels, varying graph sizes), unlike prior approaches that excel on one benchmark but degrade on others?}

    Many existing vulnerability detection methods report strong results on a single dataset but fail to generalize. We evaluate SemVul on four benchmarks with distinct properties Devign (balanced, 45\% vulnerable), Draper (imbalanced, 6\%), Reveal (imbalanced, 9\%), and PrimeVul (highly imbalanced, 3.5\%, clean labels) to demonstrate consistent effectiveness across diverse evaluation settings.
\end{itemize}

The remainder of this paper is organized as follows: Section~\ref{sec:background} reviews background and related work; Section~\ref{sec:methodology} presents our methodology; Section~\ref{sec:experimental_setup} describes the experimental setup; Section~\ref{sec:results} presents results; and Section~\ref{sec:conclusion} concludes with future directions.

  \section{Background and Related Work}
\label{sec:background}

Software vulnerability detection has evolved from manual audit and rule-based static analysis toward data-driven approaches that can scale to large codebases. Static analyzers relied on hand-crafted patterns over abstract syntax tree (AST) or control-flow graph (CFG) fragments, which delivered high precision for specific weaknesses but struggled with recall and generalization~\cite{marjanov2022mlreview}. The rise of machine learning introduced representations that encode syntactic, semantic, and control-flow context as features for classifiers. In vulnerability detection, this shift is driven by two needs: capturing long-range dependencies in code and modeling program semantics that are often invisible in surface-level token patterns~\cite{marjanov2022mlreview,chakraborty2022arewethere}. Our work follows this trajectory by using Code Property Graphs (CPGs) and multi-source embeddings to model both structure and content.
\subsection{Program Representations for Vulnerability Detection}
Program representation is central to vulnerability detection. Representations include abstract syntax trees (ASTs), control-flow graphs (CFGs), program dependence graphs (PDGs), and data-flow graphs. Each captures a different aspect of program behavior: ASTs provide syntax structure, CFGs capture execution paths, and data-flow graphs highlight variable dependencies.

Code Property Graphs unify these views into a single, typed, multi-edge graph by merging syntax, control, and data dependencies~\cite{yamaguchi2014cpg}. CPGs are a natural fit for graph learning because they expose a rich set of nodes, edges, and attributes that can be mapped to discrete vocabularies or vector embeddings. Tools such as Joern automate the extraction of CPGs from source code and have become the de facto standard in security research. In our pipeline, Joern is used to parse per-function code and export GraphSON graphs, enabling consistent downstream processing across datasets. CPGs encode different edge classes syntax edges, control-flow edges, data-dependence edges, and call relationships which can be assigned typed embeddings or one-hot encodings. Several studies show that explicitly modeling edge types improves vulnerability detection because control and data dependencies convey the flow of potentially unsafe operations~\cite{song2022hgvul,nguyen2025graph}.

A key foundation for this line of work is Devign~\cite{zhou2019devign}, which introduced a graph-based approach to data-driven vulnerability detection at the function level. Devign models source code functions as composite graphs that integrate multiple program representations abstract syntax trees, control-flow graphs, data-flow graphs, and natural code sequences into a unified structure with heterogeneous edge types. The approach employs gated graph neural networks (GGNNs) to learn node-level representations through message passing, followed by a convolutional module to aggregate graph-level features for binary classification of vulnerability status. Devign demonstrated that combining structural program analysis with deep learning significantly improves detection accuracy over purely sequential or token-based methods. This dataset has shaped subsequent research because it offers a consistent experimental setting and because its function-level focus aligns with practical code review tasks. In our study, Devign serves both as a baseline and as a reference point for the improvements introduced by CPG-driven representations and embedding fusion.

Another relevant thread in the literature compares static and dynamic analysis for vulnerability detection. Dynamic approaches can capture runtime behavior but require test inputs and impose significant overhead, which limits scalability across large repositories, whereas static approaches must approximate program behavior and can miss path-sensitive issues~\cite{marjanov2022mlreview}. CPG-based analysis provides a middle ground by encoding control and data dependencies in a static graph, allowing learning methods to reason about potential flows without executing code.

\subsection{Graph Neural Networks for Code}

Graph-based vulnerability detection builds on message passing over program graphs. Graph neural networks learn representations by iteratively aggregating information from neighbors, allowing them to encode localized and long-range dependencies. Prior work explores multiple architectures including graph convolutional networks (GCNs)~\cite{kipf2017gcn}, graph attention networks (GATs)~\cite{velickovic2018gat}, GraphSAGE~\cite{hamilton2017graphsage}, graph isomorphism networks (GINs)~\cite{xu2019gin}, gated graph neural networks (GGNNs)~\cite{li2016ggnn}, and heterogeneous or hypergraph models that handle typed edges~\cite{marjanov2022mlreview,song2022hgvul}. The Devign model itself uses a gated graph neural network to classify functions, while later work introduces hierarchical pooling methods such as DiffPool~\cite{ying2018diffpool} and SAGPool~\cite{lee2019sagpool}, edge-type attention and semantic enhancements~\cite{zhou2019devign,nguyen2025graph}.

In our architecture, the graph model is configurable with multiple GNN layers, edge-aware operators, and pooling variants such as multi-statistic pooling to capture global structure. These design choices align with the literature that emphasizes the importance of edge types, node types, and hierarchical composition in vulnerability prediction~\cite{nguyen2025graph,song2022hgvul}.

\subsection{Pre-trained Code Models}

Parallel to graph learning, transformer-based models have reshaped code understanding. Models such as CodeBERT~\cite{feng2020codebert}, GraphCodeBERT~\cite{guo2021graphcodebert}, CodeT5+~\cite{wang2023codet5plus}, and UniXcoder~\cite{guo2022unixcoder} are pretrained on large corpora with objectives tailored to code, achieving strong results in tasks like function name prediction, code search, and defect detection. These models capture token-level semantics and long-range dependencies that are difficult to express in graph-only approaches~\cite{marjanov2022mlreview}. However, they typically operate on linearized code sequences and can underutilize explicit program structure. In vulnerability detection, transformer embeddings are often used as inputs to classifiers or as features attached to graph nodes.

The vocabulary and embedding pipeline also builds on related work in representation learning. Existing code embeddings use word2vec~\cite{mikolov2013word2vec} or Code2Vec~\cite{alon2019code2vec} to capture token co-occurrence and path-based code representations, while newer approaches use subword tokenization and transformer encoders. For graph-based models, vocabularies are built for node types, edge types and content strings, and embeddings are cached to scale to large datasets.\\

\changedold{Over the last three to four years, research on automated vulnerability detection has converged on graph representations of code as its structural backbone, with the Code Property Graph (CPG) extracted by the Joern platform appearing across nearly every branch of the literature. A recent large-scale systematic review of 138 studies published between 2011 and mid-2024 confirms this picture quantitatively: graph-based input representations dominate the field at 57.2\% of the surveyed studies, deep learning models account for 88.4\% of them with graph neural networks among the most popular architectures, and Joern is the single most widely used code representation tool~\cite{harzevili2024systematic}. Building directly on this foundation, one study converts source code into CPGs, vectorises the nodes with CodeBERT, enriches them with graph centrality features, and classifies with an adaptive graph neural network that combines Transformer-style attention with graph convolutions ~\cite{liang2024source}. Going a step further, another line of research argues that language models and GNNs cover each other's blind spots and fuses them explicitly: pre-trained code language model embeddings initialise the nodes of gated GNNs over CPGs, an online knowledge distillation scheme propagates structural information across layers, and a late linear interpolation combines the two models; evaluated on four real-world datasets against seventeen state-of-the-art approaches, this fusion yields roughly 10\% F1 gains on challenging imbalanced benchmarks~\cite{liu2025vul}. The most recent studies keep the graph but move the reasoning into large language models. One such study shows that LLMs applied to plain source text overlook the syntactic and semantic structure of code, and therefore injects CPG-derived structural information into the code representation while retrieving in-context demonstrations by jointly weighing semantic, lexical, and syntactic similarity; on the Devign, Reveal, and Big-Vul datasets this approach outperforms six state-of-the-art baselines~\cite{lu2024grace}. In a similar spirit but at larger scale, a follow-up study observes that state-of-the-art detectors lose up to 45\% accuracy on rigorously verified datasets and degrade under simple code edits, and responds by using the CPG as a guide rather than an input: Joern queries extract potential vulnerable execution paths and backward slicing condenses them into focused snippets, shrinking the code shown to the LLM by 67.84\% to 90.93\% while preserving vulnerability-relevant context, which enables project-level analysis and delivers 15-40\% F1 improvements on verified datasets together with robustness to syntactic modifications~\cite{lekssays2025llmxcpg}. Taken together, these latest studies show that whether the downstream model is a GNN, a fused architecture, or an LLM, a Joern-derived graph is still computed for every function or slice, so the construction and cost of CPG-based pipelines remains shared, an important part for the entire research direction.
}\\
\changedold{While these recent works demonstrate the value of CPGs for vulnerability detection, they either use the graph structure in isolation, pair it with a single fixed semantic component, or delegate the reasoning entirely to a large language model. Our work differs in both mechanism and scope. At the mechanism level, SemVul enriches every CPG node with a precomputed semantic embedding of its code content, fused with the structural encoding before message passing, so that semantics and structure are reasoned over jointly rather than combined at the decision level; because the embeddings are precomputed, this comes without the cost of jointly training a language model. At the scope level, and to the best of our knowledge, SemVul is the first to systematically quantify the contribution of node-level semantic enrichment across multiple embedding models (from Word2Vec to CodeT5+), multiple relation-aware GNN architectures, and four benchmarks, including the rigorously labelled PrimeVul, rather than proposing
a single fixed combination. This controlled design lets us attribute performance to individual components, a question the works above leave open.}

\subsection{Hybrid Approaches}

Hybrid approaches that fuse graphs and language models are increasingly common. Some methods use transformers to encode code tokens and then map them onto graph nodes, while others perform late fusion by combining graph-level embeddings with sequence-level representations. The choice of fusion strategy affects training stability and interpretability. \changedold{In our pipeline, embeddings are extracted for both node content and edge content and merged with structural features through early concatenation at the node level, which, as we will argue in Section~\ref{sec:semantic_node}, is the fusion strategy dictated by the interdependence of semantic and structural information in this setting.} This design is motivated by prior work showing that textual content alone can miss control or data dependencies, while pure structural models can miss semantic cues~\cite{nguyen2025graph}. By integrating both, we target a representation that is robust to syntactic variation and retains contextual semantics critical for identifying vulnerabilities.

Recent work also explores self-supervised and contrastive objectives for code representations. Contrastive learning~\cite{khosla2020supcon} can align augmented views of the same function, or align graph and token representations, improving downstream classification when labeled data is limited. Approaches such as graph contrastive learning and masked token prediction aim to capture invariances that reflect program semantics rather than surface syntax. These techniques are particularly relevant in vulnerability detection, where labels are noisy and vulnerability patterns can be subtle~\cite{ji2024applying}.

\subsection{Dataset Quality and Evaluation Challenges}

Dataset preparation is a critical component in vulnerability detection research. Vulnerability datasets often suffer from duplicates, near-duplicates, or split contamination that inflates validation performance. Hash-based deduplication is a common remedy, but the hashing strategy must reflect the actual code content to avoid mismatches~\cite{chakraborty2022arewethere,chakraborty2024realvul}.

Another issue is class imbalance. Real-world code has far more non-vulnerable functions than vulnerable ones, so metrics like accuracy can be misleading. Prior work emphasizes the use of area under the ROC curve (AUC)~\cite{bradley1997auc}, precision, recall and F1-score~\cite{sokolova2009f1score} to capture detection quality~\cite{marjanov2022mlreview}. Many approaches also introduce class-balanced loss functions, focal loss variants or label smoothing to improve training stability.

An important challenge is cross-dataset generalization. Devign and Reveal are curated datasets with different labeling protocols and function extraction processes, while Draper includes broader CWE annotations and distinct code sources. Models that perform well on one dataset often degrade on another, highlighting the influence of dataset composition and preprocessing~\cite{chakraborty2024realvul}. Our pipeline is designed to minimize these confounds by standardizing preprocessing, per-function splitting, and graph extraction across datasets. This standardization enables more meaningful cross-dataset comparisons and supports claims about generalization and robustness.

Finally, another issue relates to the co-existence of different graph export formats, which expose slightly different fields. Mismatches between the graph schema and model assumptions can silently degrade performance. GraphSON, for example, contains typed nodes and edges with property dictionaries, but the mapping from properties to model features requires careful normalization. Our pipeline explicitly builds vocabularies for node types, edge types, and content strings, and maintains configuration files that mirror the extracted schema.
 \section{Methodology}
\label{sec:methodology}

This section describes our approach to semantic-enhanced vulnerability detection. We first formalize the problem, then present our code representation strategy that combines structural graphs with semantic embeddings, and finally describe our GNN architectures. Data preprocessing and implementation details are deferred to Section~\ref{sec:experimental_setup}.

\begin{figure}
\centering
\resizebox{\textwidth}{!}{\begin{tikzpicture}[
    node distance=0.6cm and 0.8cm,
    box/.style={rectangle, draw, rounded corners, minimum height=0.8cm, align=center},
    data/.style={box, fill=blue!10, minimum width=1.6cm},
    process/.style={box, fill=green!10, minimum width=1.6cm},
    storage/.style={box, fill=yellow!15, minimum width=1.4cm},
    vector/.style={box, fill=gray!10, minimum width=2.2cm, minimum height=1cm},
    concat/.style={box, fill=orange!15, thick, minimum width=2.4cm},
    arrow/.style={->, thick, >=stealth},
    label/.style={font=\itshape, text=gray},
    dimlab/.style={font=\bfseries, text=blue!70}
]

\coordinate (col0) at (0,0);
\coordinate (col1) at (3.4,0);
\coordinate (col2) at (5.8,0);
\coordinate (col3) at (8.1,0);
\coordinate (col4) at (11.0,0);
\coordinate (col5) at (13.6,0);
\coordinate (col6) at (15.4,0);
\coordinate (col7) at (18.8,0);

\node[data, minimum width=2.4cm, minimum height=1.6cm] (node) at (col0) {CPG Node\\(GraphSON)\\props: CODE, NAME};

\node[process] (gettype) at ($(col1)+(0,1.4)$) {Extract\\``label''};
\node[data] (typename) at ($(col2)+(0,1.4)$) {\texttt{CALL}};
\node[process] (onehot) at ($(col3)+(0,1.4)$) {One-Hot\\Encode};
\node[vector] (ohvec) at ($(col4)+(0,1.4)$) {
\texttt{[0,0,1,0,...,0]}\\
$T$ dimensions
};

\node[process] (gettext) at ($(col1)+(0,-1.4)$) {Extract\\``CODE''};
\node[data] (textval) at ($(col2)+(0,-1.4)$) {\texttt{strcpy}};
\node[storage] (sqlite) at ($(col3)+(0,-1.4)$) {SQLite\\Lookup};
\node[vector] (embvec) at ($(col4)+(0,-1.4)$) {
\texttt{[0.12,-0.34,...]}\\
$D$ dimensions
};

\node[concat, minimum height=1.4cm] (concat) at (col6) {
\textbf{Concatenate}\\
$[$One-Hot$|$Embedding$]$\\
Total: $T + D$ dims
};
\coordinate (concat_in_top) at ($(concat.north west)+(0,-0.2)$);
\coordinate (concat_in_bot) at ($(concat.south west)+(0,0.2)$);

\node[data, fill=red!10, minimum width=2.2cm, minimum height=1.2cm] (pygdata) at (col7) {\begin{tabular}{c}
\textbf{data.x}\\[2pt]
[N, $T$+$D$]
\end{tabular}
};

\coordinate (split) at ($(node.east)+(0.5,0)$);
\draw[arrow] (node.east) -- (split);
\draw[arrow] (split) |- (gettype.west);
\draw[arrow] (split) |- (gettext.west);
\draw[arrow] (gettype.east) -- (typename.west);
\draw[arrow] (typename.east) -- (onehot.west);
\draw[arrow] (onehot.east) -- (ohvec.west);
\draw[arrow] (gettext.east) -- (textval.west);
\draw[arrow] (textval.east) -- (sqlite.west);
\draw[arrow] (sqlite.east) -- (embvec.west);
\draw[arrow] (ohvec.east) -- ++(0.8,0) |- (concat_in_top);
\draw[arrow] (embvec.east) -- ++(0.8,0) |- (concat_in_bot);
\draw[arrow] (concat.east) -- (pygdata.west);

\node[dimlab, above=0.15cm of gettype] {Path A: Structural};
\node[dimlab, below=0.15cm of gettext] {Path B: Semantic};

\node[box, fill=yellow!20, minimum width=6cm, align=left, below=1.1cm of concat, xshift=-1cm] (legend) {
\textbf{Dimensions:} $T$ = \#node types (dataset-specific: Devign=15, Reveal/Draper=16)\\
$D$ = embedding dim: Word2Vec (100--200), CodeBERT/UniXcoder (768), CodeT5+ (256)
};

\end{tikzpicture}}
\caption{Feature construction pipeline for each CPG node. \textbf{Path A (Structural)}: Node type (e.g., CALL, IDENTIFIER) is converted to a sparse one-hot encoding ($T$ dimensions, where $T$ is dataset-specific: Devign=15, Reveal=16, Draper=16). \textbf{Path B (Semantic)}: Node content (source code text) is normalized, hashed with SHA256 for efficient lookup, and the pre-computed embedding ($D$ dimensions) is retrieved from SQLite cache. Both vectors are concatenated to form the final node feature vector ($T$+$D$ dimensions). Embedding dimension $D$ varies by model: Word2Vec (100-200), CodeBERT/UniXcoder (768), CodeT5+ (256).}
\label{fig:feature_construction}
\end{figure}
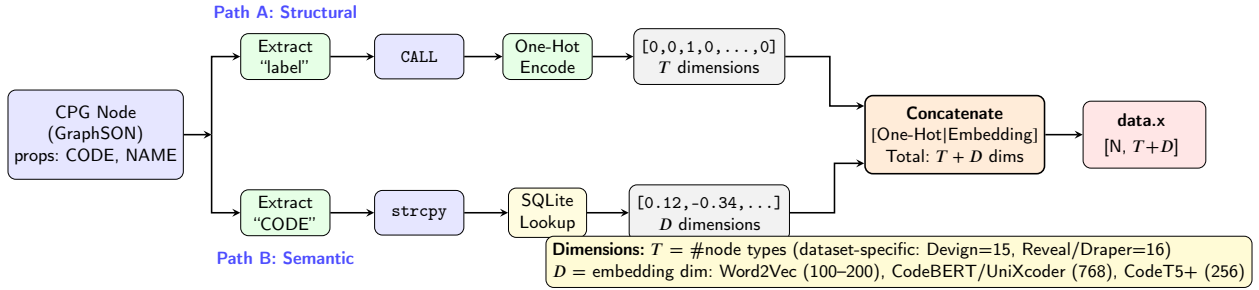

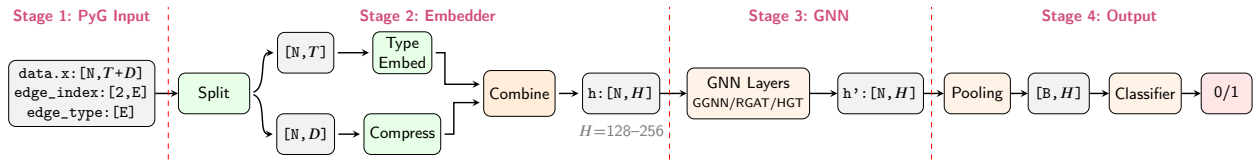
\begin{figure}
\centering
\resizebox{\textwidth}{!}{\begin{tikzpicture}[
    node distance=0.5cm and 0.6cm,
    box/.style={rectangle, draw, rounded corners, minimum height=0.8cm, align=center},
    data/.style={box, fill=blue!10},
    process/.style={box, fill=green!10},
    model/.style={box, fill=orange!10},
    output/.style={box, fill=red!10},
    tensor/.style={box, fill=gray!10, font=\small\ttfamily, minimum width=1.5cm},
    arrow/.style={->, thick, >=stealth},
    label/.style={font=\small\itshape, text=gray},
    stage/.style={font=\small\bfseries, text=purple!70}
]

\coordinate (c0) at (-1,0);
\coordinate (c1) at (1.6,0);
\coordinate (c2) at (3.4,0);
\coordinate (c3) at (5.3,0);
\coordinate (c4) at (7.6,0);
\coordinate (c5) at (9.6,0);
\coordinate (c6) at (12.1,0);
\coordinate (c7) at (14.7,0);
\coordinate (c8) at (16.6,0);
\coordinate (c9) at (18.2,0);
\coordinate (c10) at (19.9,0);
\coordinate (c11) at (21.5,0);

\node[stage] at ($(c0)+(0,2.3)$) {Stage 1: PyG Input};
\node[tensor, minimum width=2cm, minimum height=1.3cm] (input) at ($(c0)+(0,0.8)$) {
\textbf{data.x}:[N,$T$+$D$]\\
\textbf{edge\_index}:[2,E]\\
\textbf{edge\_type}:[E]
};

\node[stage] at ($(c3)+(0.5,2.3)$) {Stage 2: Embedder};
\node[process, minimum width=1.4cm] (split) at ($(c1)+(0,0.8)$) {Split};
\node[tensor, minimum width=0.8cm] (oh) at ($(c2)+(0,1.6)$) {[N,$T$]};
\node[tensor, minimum width=0.8cm] (emb) at ($(c2)+(0,0.0)$) {[N,$D$]};
\node[process, minimum width=1.1cm] (typeproj) at ($(c3)+(0,1.6)$) {Type\\Embed};
\node[process, minimum width=1.1cm] (compress) at ($(c3)+(0,0.0)$) {Compress};
\node[process, fill=orange!15, minimum width=1.3cm, minimum height=0.9cm] (concatproj) at ($(c4)+(0,0.8)$) {Combine};
\coordinate (cp_in_top) at ($(concatproj.west)+(0,0.18)$);
\coordinate (cp_in_bot) at ($(concatproj.west)+(0,-0.18)$);
\node[tensor, minimum width=1.4cm] (hidden) at ($(c5)+(0,0.8)$) {\textbf{h}:[N,$H$]};

\node[stage] at ($(c6)+(1,2.3)$) {Stage 3: GNN};
\node[model, minimum width=1.8cm, minimum height=1.0cm] (gnn) at ($(c6)+(0,0.8)$) {
GNN Layers\\
{\scriptsize GGNN/RGAT/HGT}
};
\node[tensor, minimum width=1.5cm] (gnnout) at ($(c7)+(0,0.8)$) {\textbf{h'}:[N,$H$]};

\node[stage] at ($(c9)+(0.8,2.3)$) {Stage 4: Output};
\node[model, minimum width=1.3cm] (pool) at ($(c8)+(0,0.8)$) {Pooling};
\node[tensor, minimum width=1.2cm] (graphemb) at ($(c9)+(0,0.8)$) {[B,$H$]};
\node[model, minimum width=1.2cm] (mlp) at ($(c10)+(0,0.8)$) {Classifier};
\node[output, minimum width=1.0cm] (pred) at ($(c11)+(0,0.8)$) {0/1};

\draw[arrow] (input.east) -- (split.west);
\draw[arrow, shorten >=2pt, shorten <=2pt] (split.east) to[out=20,in=180] (oh.west);
\draw[arrow, shorten >=2pt, shorten <=2pt] (split.east) to[out=-20,in=180] (emb.west);
\draw[arrow, shorten >=2pt, shorten <=2pt] (oh.east) -- (typeproj.west);
\draw[arrow, shorten >=2pt, shorten <=2pt] (emb.east) -- (compress.west);
\draw[arrow, shorten >=2pt, shorten <=2pt] (typeproj.east) -- ++(0.25,0) |- (cp_in_top);
\draw[arrow, shorten >=2pt, shorten <=2pt] (compress.east) -- ++(0.25,0) |- (cp_in_bot);
\draw[arrow, shorten >=2pt, shorten <=2pt] (concatproj.east) -- (hidden.west);

\draw[arrow] (hidden.east) -- (gnn.west);

\draw[arrow] (gnn.east) -- (gnnout.west);
\draw[arrow] (gnnout.east) -- (pool.west);
\draw[arrow] (pool.east) -- (graphemb.west);
\draw[arrow] (graphemb.east) -- (mlp.west);
\draw[arrow] (mlp.east) -- (pred.west);

\node[text=gray, below=0.1cm of hidden] {$H$=128--256};

\draw [dashed, color=red] (0.7,-0.5) -- (0.7,2.5);
\draw [dashed, color=red] (10.55,-0.5) -- (10.55,2.5);
\draw [dashed, color=red] (15.7,-0.5) -- (15.7,2.5);

\end{tikzpicture}}
\caption{Complete data flow from input to prediction. \textbf{Stage 1}: PyG Data contains node features \texttt{data.x} of shape [N, $T$+$D$] (one-hot type encoding concatenated with semantic embedding), graph structure \texttt{data.edge\_index} of shape [2, E] where the two rows store source and destination node indices respectively for each of the $E$ edges (COO sparse format), and relation types \texttt{data.edge\_type} of shape [E] storing the edge type index for each edge. \textbf{Stage 2}: The embedder \emph{splits} the concatenated input back into its one-hot block ($T$ dims) and semantic block ($D$ dims). The one-hot block is converted to learned dense embeddings, while the semantic block is compressed to a lower dimension. Both outputs are concatenated and passed through a linear layer to produce unified hidden representations [N, $H$]. \textbf{Stage 3}: GNN performs $L$ rounds of edge-type-aware message passing using relation-specific weights $\mathbf{W}_r$, producing updated node embeddings $\mathbf{h'}$. \textbf{Stage 4}: Pooling (e.g., mean, max, attention, or multi-stat) aggregates per-node embeddings to a single graph-level vector [B, $H$], and the classifier (a two-layer MLP with LayerNorm, GELU, and dropout) maps this to a binary vulnerability prediction. Symbols: $N$ = nodes per graph, $E$ = edges per graph, $B$ = graphs per batch, $T$ = node-type dimensions, $D$ = embedding dimensions, $H$ = hidden dimensions.}
\label{fig:complete_dataflow}
\end{figure}

Figure~\ref{fig:system_architecture} summarizes our complete pipeline: source code is parsed into CPGs, enriched with semantic embeddings, processed by GNNs that exploit edge-type heterogeneity, pooled to graph-level representations, and classified. This design addresses the limitations of prior work by combining structural program understanding with semantic code meaning. Figure~\ref{fig:feature_construction} and Figure~\ref{fig:complete_dataflow} provide a more detailed depiction of our pipeline, showing respectively the initial feature construction executed per each CPG node, and the subsequent data flow leading to the predictions.

\changedold{Returning to the running example, Figure~\ref{fig:feature_construction} shows how
the \texttt{CALL strcpy} node of Figure~\ref{fig:code_to_graph} becomes a
feature vector. Path~A extracts its label \texttt{CALL} and one-hot encodes it
into $T$ dimensions ($T{=}15$ for Devign), a vector that would be identical
for a call to \texttt{strncpy}. Path~B extracts its code content
\texttt{strcpy}, normalises and hashes it, and retrieves its pre-computed
$D$-dimensional semantic embedding from the Embedder (e.g., $D{=}768$ for
GraphCodeBERT); this is the component that separates \texttt{strcpy} from
\texttt{strncpy}. The two blocks are concatenated into the final $T{+}D$
node feature, and the same procedure is applied to the remaining six nodes,
yielding the matrix \texttt{data.x} of shape $[7,\,T{+}D]$.
}

\changedold{Figure~\ref{fig:complete_dataflow} completes the running example by following the
graph of Figure~\ref{fig:feature_construction} to a prediction. In Stage~1 the
function is a PyG object with \texttt{data.x} of shape $[7,\,T{+}D]$, an
\texttt{edge\_index} listing the source and destination of every AST, CFG, and
DFG edge in Figure~\ref{fig:code_to_graph}, and an \texttt{edge\_type} vector
recording each edge's relation. Stage~2 splits each node feature back into its
one-hot and semantic blocks and maps them to a unified $H$-dimensional hidden
state. In Stage~3, edge-type-aware message passing propagates information
along the typed edges: through the relation-specific weight of the DFG
relation, the \texttt{CALL strcpy} node receives the representation of
\texttt{PARAM input} via \texttt{ARG input}, while no bounds-checking node
exists on the control-flow path to attenuate it; after $L$ layers the call
node therefore encodes the joint pattern ``risky API receiving unchecked,
attacker-reachable data''. Stage~4 pools the node embeddings into one
graph vector and the MLP classifier maps it to the positive (vulnerable)
class.
}

\subsection{Problem Formulation}

We formulate vulnerability detection as a binary graph classification problem. Given a function $f$ represented as a Code Property Graph $G = (V, E, X, \mathcal{T}_v, \mathcal{T}_e)$, where $V = \{v_1, \ldots, v_n\}$ is the set of nodes representing code elements, $E \subseteq V \times V$ is the set of directed edges representing relationships, $X \in \mathbb{R}^{n \times d}$ is the node feature matrix, $\mathcal{T}_v: V \rightarrow \{1, \ldots, K\}$ maps nodes to $K$ node types, and $\mathcal{T}_e: E \rightarrow \{1, \ldots, R\}$ maps edges to $R$ relation types. 
The goal is to learn a function $\Phi: G \rightarrow \{0, 1\}$ that predicts whether the function contains a vulnerability.

\subsection{Code Property Graph Representation}

We use Joern~\cite{yamaguchi2014cpg} to generate Code Property Graphs that unify multiple program representations into a single heterogeneous graph:
\begin{equation}
    G_{\text{CPG}} = G_{\text{AST}} \cup G_{\text{CFG}} \cup G_{\text{DFG}} \cup G_{\text{PDG}}
\end{equation}

\noindent where $G_{\text{AST}}$ captures syntactic structure, $G_{\text{CFG}}$ captures execution order, $G_{\text{DFG}}$ captures data dependencies, and $G_{\text{PDG}}$ captures program dependencies.

CPG nodes represent different code elements, each with distinct roles in vulnerability patterns. The source code text for each node is extracted from the CPG's \texttt{CODE} property, which we later transform into semantic embeddings (see Section~\ref{sec:semantic_node}). Key node categories include \textbf{Function Call Nodes} (invocations of functions, the most vulnerability-critical node type since dangerous functions like \texttt{strcpy}, \texttt{sprintf}, and \texttt{gets} are represented as call nodes),
\textbf{Identifier Nodes} (variable references that track data flow from tainted sources),
\textbf{Literal Nodes} (constant values including buffer sizes, e.g., \texttt{char buf[10]}),
\textbf{Control Structure Nodes} (conditionals and loops whose presence (or absence) determines whether safety checks exist),
\textbf{Declaration Nodes} (local variable declarations including buffer allocations), and
\textbf{Entry/Exit Nodes} (function entry and return points, relevant for resource leak detection and scope analysis). The node type vocabulary varies across datasets: Devign has 15 node types while Reveal and Draper have 16 (including struct member access patterns). This schema heterogeneity means models must adapt their type encoding dimensions per dataset. The complete schema for each dataset is provided in Appendix~\ref{app:schemas}.

The CPG edge types carry distinct semantic meanings critical for vulnerability detection. Each edge category corresponds to one of the underlying program representations (AST, CFG, DFG, PDG) introduced in Section~\ref{sec:background}.
\textbf{Data-Flow Edges} (from $G_{\text{DFG}}$) show where a variable's value reaches its use, critical for tracking tainted data from sources (e.g., user input) to sinks (e.g., unsafe copy functions).
\textbf{Control-Flow Edges} (from $G_{\text{CFG}}$) encode execution order between statements, and are thus essential for detecting missing bounds checks or null guards.
\textbf{Syntax Edges} (from $G_{\text{AST}}$) represent parent-child relationships in the abstract syntax tree, capturing code structure but less directly vulnerability-relevant.
\textbf{Control-Dependence Edges} (from $G_{\text{PDG}}$) show which statements are conditionally executed, thus are important for understanding branch coverage of safety checks.
\textbf{Dominance Edges} represent relationships in control flow identifying mandatory execution paths.
Finally, \textbf{Call-Related Edges} encode function call and argument relationships, and are critical since dangerous library functions are often sites of vulnerabilities. Different vulnerability types manifest through different edge patterns: buffer overflows appear in data-flow chains to unsafe functions, while null pointer dereferences appear as control-flow paths lacking protective conditionals.

CPGs provide three key advantages over token sequences. First, vulnerabilities often involve data flowing from untrusted sources to sensitive sinks. Data-flow edges explicitly encode these dependencies. Second, missing bounds checks or null pointer guards are visible in control-flow structure the path to a dangerous operation lacks a protective branch. Third, the graph topology remains stable under variable renaming, comment changes, or whitespace modifications that confuse token-based models. Figure~\ref{fig:code_to_graph} illustrates this transformation for a buffer overflow vulnerability. The CPG captures how untrusted input flows through the program structure to a dangerous function call information that is invisible to token-based models.

\subsection{Semantic-Enhanced Node Representation}
\label{sec:semantic_node}

A key contribution of SemVul is enriching CPG nodes with semantic embeddings. Each node $v_i$ is represented by $\mathbf{x}_i = [\mathbf{t}_i \| \mathbf{e}_i]$
, where $\mathbf{t}_i \in \{0,1\}^K$ is a one-hot encoding of the node type (one of $K$ categories such as function call, identifier, or literal), and $\mathbf{e}_i \in \mathbb{R}^{d_e}$ is a semantic embedding of the node's code content.

We explore multiple embedding sources. First, we train Word2Vec models~\cite{mikolov2013word2vec} on the code vocabulary extracted from each dataset. Tokens are split on CamelCase and snake\_case boundaries, and embeddings are aggregated via mean pooling. Then, for richer semantic understanding, we use embeddings from CodeBERT~\cite{feng2020codebert}, GraphCodeBERT~\cite{guo2021graphcodebert}, UniXcoder~\cite{guo2022unixcoder}, and CodeT5+~\cite{wang2023codet5plus}. \changedold{These models are pre-trained on large code corpora with general-purpose objectives, i.e., they are not specifically trained for vulnerability detection purposes. We therefore hypothesize that the semantic representations they produce also encode contextual patterns correlated with vulnerability. This is an expectation that must be verified empirically, which is precisely the aim of RQ1.}

\changed{The pre-trained code models are frozen and are never fine-tuned. Each model is applied in a single forward pass (inference mode) to precompute embeddings for the code content of the CPG nodes. The resulting vectors are cached and later loaded as fixed node features, so the language model itself is not part of the trained network and is not even loaded during training. 
The same holds for Word2Vec, with one difference in provenance. Following the original Devign protocol, we train it ourselves on the code tokens of each dataset, but using only the standard unsupervised objective and without any vulnerability labels, after which it is likewise frozen and used only to look up vectors.
Because the embeddings enter the model as fixed inputs rather than trainable parameters, they receive no gradient at any point. The reported semantic enhancement therefore reflects the information already present in these representations, and not any task-specific adaptation of the underlying language models.}

\changedold{The semantics has to be seen as a complement to the structure, rather than as an independent source of information: identical structural patterns may correspond to entirely different vulnerability behavior depending on their semantic content, and conversely the semantics of a code fragment is only meaningful in relation to the structure it belongs to. This logical correlation corresponds to the early fusion paradigm, in which different measures are stacked together to create the feature vector. Late fusion, in contrast, is appropriate when separate sources of information can each independently support the decision, which is not the case here: no decision on the vulnerability of the code could be taken from the semantics alone. For this reason, node-level semantic embeddings are concatenated with the structural one-hot type encoding to form a single node feature vector.}

\changedold{Following, we use an example to explain intuitively why semantic embeddings
complement structure. Consider a node representing the call
\texttt{strcpy(buffer, input)}. With one-hot encoding alone, this node is
indistinguishable from any other CALL node, \texttt{strcpy} and
\texttt{strncpy} receive identical feature vectors, so no downstream model can
treat them differently. Semantic embeddings remove this bottleneck in three
steps. First, pre-trained models such as CodeBERT make the two calls
\emph{distinguishable} trained with masked-token objectives on large corpora,
they assign different vectors to tokens used in different contexts, and
\texttt{strncpy} systematically co-occurs with length arguments and
\texttt{sizeof} expressions while \texttt{strcpy} does not. We stress that the
embedding encodes no notion of safety; it only guarantees that the classifier
receives different inputs for the two calls. Second, the \emph{association
with vulnerability is learned during supervised training}: since vulnerable
functions contain unchecked \texttt{strcpy}-like calls far more often, the GNN
learns to map that region of the embedding space to higher risk. Third, the
\emph{CPG structure determines whether the risk is realised}: data-flow edges
reveal whether \texttt{input} originates from an unvalidated source, and
control-flow edges reveal whether a bounds check precedes the call, so message
passing combines the call's semantics with its context and the model learns
the joint pattern ``risky API with unchecked input'' rather than the mere
presence of a token. This also explains our early-fusion choice. The decision
emerges from semantics and structure jointly, and neither is alone sufficient. 
}

\subsection{Graph Neural Network Architectures}

All our GNN architectures follow the Message Passing Neural Network framework~\cite{gilmer2017mpnn}, where each node iteratively gathers messages from its neighbors, aggregates them using a permutation-invariant function (sum, mean, or max), and updates its representation. After $L$ layers, each node encodes information from its $L$-hop neighborhood. We evaluate four architectures. 
\changedold{The selection of the four architectures was driven by the nature of our input data rather than by a desire for architectural variety. A Code Property Graph is not a plain graph: it is a multi-relational structure in which nodes carry code semantics and edges belong to distinct types (i.e., AST, control-flow, data-flow), and each edge type carries different information and, mathematically, deserves a different transformation or weight during message passing. This immediately rules out architectures that treat all edges as equivalent, and it defines the design space we sampled from: models that differ in how they exploit typed-edge structure.
Within that space, each of the four architectures represents one distinct and well-motivated hypothesis about how edge-type information should be handled. GGNN assigns each edge type its own weight matrix and uses a gated (GRU) update, testing whether type-specific transformations plus long-range memory are sufficient. RGAT adds learned attention per edge type, testing whether the model benefits from deciding that some relations (for example data-flow when tracking a tainted value) matter more than others for a given node. Multi-View GGNN takes the opposite decomposition: instead of mixing all edge types in one pass, it processes the AST, CFG, and DFG as separate views with separate encoders and fuses them by attention, testing whether vulnerabilities are better captured when each program aspect is modelled independently. Finally, the Call-Aware GNN encodes a domain-specific prior, namely that vulnerabilities frequently manifest at call sites of dangerous functions, and gives CALL nodes and their data-flow context special treatment.
Importantly, all four share the same message-passing framework, the same input representation, and the same pooling and classification head, so the comparison is controlled, the only thing that varies is the inductive bias about edge types. Our goal in evaluating them was therefore not to increase model complexity, but to determine empirically which way of handling multi-relational code graphs handles our data best, and to report that comparison rather than a single-architecture result.}

\textbf{Gated Graph Neural Networks (GGNN)}~\cite{li2016ggnn} employ edge-type-specific weight matrices to transform neighbor representations differently based on relationship type (e.g., data-flow vs.\ control-flow edges). The update mechanism uses a Gated Recurrent Unit (GRU), a recurrent neural network component that selectively remembers or forgets information across layers, which is crucial for capturing long-range dependencies such as when a variable defined at the start of a function is used unsafely much later.

\textbf{Relational Graph Attention Networks (RGAT)}~\cite{velickovic2018gat,schlichtkrull2018rgcn} combine relational modeling with attention mechanisms. The key insight is that not all neighbors are equally important, and different edge types carry different significance. Attention coefficients are learned per edge type, allowing the model to focus heavily on data-flow edges for tracking tainted values while attending to control-flow edges for detecting missing safety checks.

\textbf{Multi-View GGNN}~\cite{zhuang2021software} decompose the CPG into separate views (AST, CFG, DFG) and processes each independently using separate GGNN encoders, rather than processing all edge types together. The view-specific representations are then combined via learned attention weights, which is particularly effective when different vulnerability types manifest in different program aspects buffer overflows in data flow, missing checks in control flow.

\textbf{Call-Aware GNN}~\cite{cheng2021deepwukong} are based on the observation that function calls are where vulnerabilities often manifest. This architecture applies special attention to CALL nodes and explicitly models their data-flow context by aggregating incoming and outgoing data-flow edges. This design is motivated by the fact that dangerous functions like \texttt{strcpy}, \texttt{memcpy}, and \texttt{sprintf} are ``choke points'' where improper inputs cause vulnerabilities.

\subsection{Graph-Level Pooling and Classification}

After message passing, we aggregate node representations into a graph-level embedding for classification. Our multi-statistic pooling~\cite{ying2018diffpool} concatenates mean, max, and standard deviation over all node representations, capturing average node behavior, extreme signals that may indicate anomalies, and variability across the graph respectively. This richer distributional representation outperforms single-statistic approaches (e.g., mean-only or max-only pooling) by 3+ F1 points in our experiments. The pooled representation is then passed through a two-layer MLP (multilayer perceptron) classifier with ReLU (Rectified Linear Unit) activation function and sigmoid output for binary vulnerability prediction.
We use weighted binary cross-entropy loss with class weights inversely proportional to class frequencies to handle the severe class imbalance present in vulnerability datasets.

 \section{Experimental Setup}
\label{sec:experimental_setup}

This section describes our experimental configuration, including the datasets used, model hyperparameters, and evaluation metrics. All experiments were conducted on an NVIDIA RTX 5000 Ada Generation GPU (32GB VRAM) with CUDA 13.0 and driver version 580.95.05. Our implementation uses PyTorch 2.0 with PyTorch Geometric~\cite{fey2019pytorchgeometric}, Joern version 4.0.399 for CPG extraction. To improve training efficiency, we pre-computed and cached both one-hot node type encodings and semantic embeddings, enabling fast graph construction and data loading during training. We used mixed precision training (FP16) where supported to reduce memory usage. Training times vary by architecture: using EdgeType-GNN as baseline, RGAT requires approximately 1.5$\times$ longer due to attention computation, while HGT requires approximately 3$\times$ longer due to heterogeneous transformer layers; all experiments completed within 24 hours per dataset. Random seeds were fixed at 42 for reproducibility. Class imbalance was handled using weighted loss functions with $\text{pos\_weight} = n_{\text{neg}}/n_{\text{pos}}$.

\subsection{Datasets}
\label{sec:datasets}

We evaluate our approach on four widely-used vulnerability detection benchmarks. Each dataset contains C/C++ functions labeled as vulnerable or safe. Table~\ref{tab:datasets} summarizes the key statistics. \textbf{Devign} is the most commonly used benchmark for graph-based vulnerability detection. It contains functions extracted from FFmpeg and Qemu with expert-verified labels. \textbf{Draper} provides a larger and more diverse collection with CWE annotations. \textbf{Reveal} was specifically designed to address data quality issues in earlier datasets by using careful labeling procedures. \textbf{PrimeVul}~\cite{ding2024primevul} is a recent high-quality benchmark with rigorous labeling that addresses the label noise problems in earlier datasets; it is highly imbalanced (3.5\% vulnerable), and thus a challenging evaluation setting.

\begin{table}[t]
\centering
\caption{Dataset Statistics}
\label{tab:datasets}

\begin{tabular}{lrrrrr}
\hline
\textbf{Dataset} & \textbf{Funcs} & \textbf{Nodes} & \textbf{Edges} & \textbf{NT} & \textbf{ET} \\
\hline
Devign & 27,318 & 9.65M & 68.2M & 15 & 18 \\
Draper & 127,437 & 24.1M & 137.3M & 16 & 19 \\
Reveal & 22,383 & 5.69M & 38.0M & 16 & 19 \\
PrimeVul & 235,021 & 42.3M & 245.8M & 16 & 19 \\
\hline
\multicolumn{6}{l}{\small NT = Node Types, ET = Edge Types}
\end{tabular}
\end{table}

Table~\ref{tab:per_function_stats} shows per-function graph statistics. The variation in graph sizes across datasets reflects differences in function complexity and coding styles. Table~\ref{tab:split_stats} provides detailed per-split statistics including class distribution, which is critical for understanding dataset imbalance. Looking at the statistics, we observe that 
\textbf{Devign is nearly balanced} ($\sim$45\% vulnerable), making standard metrics meaningful, while
\textbf{Draper and Reveal are moderately imbalanced} ($\sim$6\% and $\sim$9\% vulnerable respectively), requiring weighted loss functions and careful metric interpretation, and 
\textbf{PrimeVul is extremely imbalanced} ($\sim$3.5\% vulnerable) with high-quality labels, representing the most challenging and realistic evaluation setting. Also, we observe that \textbf{graph sizes vary significantly}: Devign has functions with up to 12,294 nodes, while Reveal reaches 18,682 nodes.

\begin{table}[t]
\centering
\caption{Per-Function Graph Statistics (from Cache Index)}
\label{tab:per_function_stats}
\begin{tabular}{@{}lrrrr@{}}
\toprule
\textbf{Dataset} & \textbf{Avg. N} & \textbf{Avg. E} & \textbf{Max N} & \textbf{Max E} \\
\midrule
Devign & 243.0 & 1,799.0 & 12,294 & 109,293 \\
Draper & 189 & 1,078 & 958 & 11,598 \\
Reveal & 167.9 & 1,165.5 & 18,682 & 124,621 \\
PrimeVul & 180.2 & 1,045.8 & 15,421 & 98,347 \\
\bottomrule
\end{tabular}
\end{table}

\begin{table}[t]
\centering
\caption{Per-Split Statistics and Class Imbalance}
\label{tab:split_stats}
\begin{tabular}{@{}llrrr@{}}
\toprule
\textbf{Dataset} & \textbf{Split} & \textbf{Funcs} & \textbf{Vuln.} & \textbf{\%} \\
\midrule
Devign & Train & 21,854 & 10,018 & 45.8 \\
 & Val & 2,732 & 1,187 & 43.4 \\
 & Test & 2,732 & 1,255 & 45.9 \\
\midrule
Draper & Train & 101,950 & 6,321 & 6.2 \\
 & Val & 12,744 & 812 & 6.4 \\
 & Test & 12,743 & 785 & 6.2 \\
\midrule
Reveal & Train & 17,974 & 1,697 & 9.4 \\
 & Val & 2,200 & 175 & 8.0 \\
 & Test & 2,209 & 196 & 8.9 \\
\midrule
PrimeVul & Train & 188,017 & 6,580 & \textbf{3.5} \\
 & Val & 23,502 & 823 & 3.5 \\
 & Test & 23,502 & 822 & 3.5 \\
\bottomrule
\end{tabular}
\end{table}

\subsection{Data Preprocessing}
\label{sec:data_preprocessing}

Reliable vulnerability detection requires careful data engineering to avoid representation mismatch, split contamination, and dataset artifacts. This section describes our preprocessing pipeline.

All datasets are converted into a unified format: each split (train/validation/test) is stored as a directory of per-function C files, associate with a CSV that records function ID, vulnerability label, and content hash. Joern processes each function to produce CPG exports in GraphSON format. This standardization allows us to use the same training code across datasets and ensures that performance differences reflect dataset content rather than preprocessing variations.

CPG extraction produces \emph{different} node and edge type vocabularies depending on the dataset, code patterns, and Joern version. This heterogeneity is a critical observation: \textbf{different datasets cannot be treated identically}, as their underlying graph schemas differ. Our pipeline explicitly discovers and validates these schemas. Table~\ref{tab:schema_summary} summarizes the schema dimensions for each dataset. The complete list of node and edge types with their descriptions is provided in Appendix~\ref{app:schemas}. This schema heterogeneity has important implications for model design: one-hot dimensions differ across datasets (Devign uses 15-dimensional node type encodings while Reveal/Draper/PrimeVul use 16-dimensional encodings), edge-type-aware GNNs must adapt their weight matrices accordingly, and cross-dataset transfer requires vocabulary mapping and retraining. Our vocabulary builder explicitly discovers these schemas from GraphSON exports and produces synchronized vocabulary files that ensure consistent index assignments within each dataset, preventing silent degradation from missing edge categories or schema drift.

\begin{table}[t]
\centering
\caption{CPG Schema Summary by Dataset}
\label{tab:schema_summary}
\begin{tabular}{@{}lccc@{}}
\toprule
\textbf{Dataset} & \textbf{NT} & \textbf{ET} & \textbf{Notable Additions} \\
\midrule
Devign & 15 & 18 & Base schema \\
Reveal & 16 & 19 & MEMBER, CAPTURE \\
Draper & 16 & 19 & Similar to Reveal \\
PrimeVul & 16 & 19 & Similar to Reveal \\
\bottomrule
\hline
\multicolumn{4}{l}{\small NT = Node Types, ET = Edge Types}
\end{tabular}
\end{table}

To enable efficient experimentation with different embedding strategies, we pre-compute embeddings for all unique node content strings, and store them in SQLite databases indexed by content hash, enabling fast lookup during dataset construction. 
In particular, \textbf{Word2Vec} is trained per dataset on the extracted code vocabulary, while 
\textbf{pre-trained Models} (CodeBERT, GraphCodeBERT, UniXcoder, CodeT5+) embeddings are computed once and cached.
Cache signatures encode the embedding configuration to prevent mismatches.

The final datasets are constructed by loading GraphSON exports for each function, mapping node and edge types to vocabulary indices, concatenating one-hot type encodings with semantic embeddings, and constructing sparse adjacency representations with edge type labels. The resulting cached datasets support reproducible training across embedding and architecture configurations.

\subsection{Model Configurations}
\label{sec:model_configs}

Table~\ref{tab:model_configs} summarizes the configurations for each model variant. We tested both simple baselines (like MLP without GNN) and sophisticated architectures (like RGAT and HGT) to understand the contribution of different components.

\begin{table}[t]
\centering
\caption{Model Configurations}
\label{tab:model_configs}
\begin{tabular}{@{}lcccc@{}}
\toprule
\textbf{Model} & \textbf{GNN} & \textbf{Layers} & \textbf{Pool} & \textbf{Embed.} \\
\midrule
Devign & GGNN & 4 & Conv1D & learned \\
Improved & GGNN & 1--2  & multi\_stat & word2vec \\
Call-Aware & call\_aware & 4  & multi\_stat & word2vec \\
MLP & none & 0  & multi\_stat & codebert \\
RGAT & rgat & 4 & Conv1D & learned \\
MultiView & ggnn & 4 &  Conv1D & learned \\
HGT & hgt & 3 & Conv1D & learned \\
\bottomrule
\hline
\multicolumn{5}{l}{\small Hidden dimension = $T{+}D$  for all models}
\end{tabular}
\end{table}

We experimented with multiple code embedding strategies to understand their impact on detection performance. Table~\ref{tab:embeddings} shows the embedding configurations used in our experiments.

\begin{table}[t]
\centering
\caption{Node Embedding Configurations}
\label{tab:embeddings}
\begin{tabular}{lccl}
\toprule
\textbf{Embedder} & \textbf{Dim} & \textbf{Pre-train} & \textbf{Description} \\
\midrule
One-Hot Only & 16 & No & Type encoding only \\
Word2Vec-100d & 116 & No & Per-dataset trained \\
Word2Vec-200d & 216 & No & Per-dataset trained \\
CodeT5+ & 272 & Yes & 16 + 256 dims \\
CodeBERT & 784 & Yes & 16 + 768 dims \\
GraphCodeBERT & 784 & Yes & 16 + 768 dims \\
UniXcoder & 784 & Yes & 16 + 768 dims \\
\bottomrule
\end{tabular}
\end{table}

For all models, we set the hidden dimension to the total input feature size ($T{+}D$), where $T$ is the node-type one-hot size and $D$ is the embedding dimension. For example:
\begin{itemize}
    \item Devign + Word2Vec-200d uses $T{+}D=15{+}200=215$;
    \item Reveal/Draper/PrimeVul + Word2Vec-200d uses $16{+}200=216$;
    \item CodeT5+ uses $16{+}256=272$;
    \item CodeBERT/GraphCodeBERT/UniXcoder use $16{+}768=784$.
\end{itemize}
All models were trained using the same optimization setup to ensure fair comparison. Table~\ref{tab:training_hyperparams} lists the training hyperparameters.

\begin{table}[t]
\centering
\caption{Training Hyperparameters}
\label{tab:training_hyperparams}
\begin{tabular}{lc}
\hline
\textbf{Hyperparameter} & \textbf{Value} \\
\hline
Optimizer & AdamW \\
Learning Rate & 0.0005 \\
Weight Decay & 0.01 \\
Batch Size & 64 \\
Maximum Epochs & 100 (all models) \\
Early Stopping Patience & 30 \\
Dropout Rate & 0.2 \\
Gradient Clipping & 1.0 \\
LR Scheduler & ReduceLROnPlateau \\
Scheduler Patience & 10 \\
\hline
\end{tabular}
\end{table}

\changedold{
\subsection{Evaluation Metrics}
\label{sec:metrics}
We evaluate vulnerability detection effectiveness using five standard
classification metrics~\citep{sokolova2009f1score,bradley1997auc}. All
metrics are derived from the four entries of the confusion matrix computed on
the test set: true positives $\mathit{TP}$ (vulnerable functions correctly
identified), true negatives $\mathit{TN}$ (safe functions correctly identified),
false positives $\mathit{FP}$ (safe functions flagged as vulnerable), and false
negatives $\mathit{FN}$ (vulnerable functions missed).

\emph{Accuracy} measures the overall fraction of correctly classified functions:
\begin{equation*}
\mathrm{Accuracy} = \frac{\mathit{TP} + \mathit{TN}}
{\mathit{TP} + \mathit{TN} + \mathit{FP} + \mathit{FN}}.
\end{equation*}

\emph{Precision} is the fraction of functions flagged as vulnerable that are
truly vulnerable, and \emph{Recall} is
the fraction of vulnerable functions that are correctly identified:
\begin{equation*}
\mathrm{Precision} = \frac{\mathit{TP}}{\mathit{TP} + \mathit{FP}},
\qquad
\mathrm{Recall} = \frac{\mathit{TP}}{\mathit{TP} + \mathit{FN}}.
\end{equation*}

\emph{F1 score} is the harmonic mean of precision and recall, balancing false
positives against false negatives:
\begin{equation*}
\mathrm{F1} = 2 \cdot
\frac{\mathrm{Precision} \cdot \mathrm{Recall}}
{\mathrm{Precision} + \mathrm{Recall}}
= \frac{2\,\mathit{TP}}{2\,\mathit{TP} + \mathit{FP} + \mathit{FN}}.
\end{equation*}

\emph{AUC} is the area under the ROC curve, which plots the true positive rate
$\mathrm{TPR} = \mathit{TP}/(\mathit{TP}+\mathit{FN})$ against the false positive
rate $\mathrm{FPR} = \mathit{FP}/(\mathit{FP}+\mathit{TN})$ as the classification
threshold varies:
\begin{equation*}
\mathrm{AUC} = \int_{0}^{1} \mathrm{TPR}\bigl(\mathrm{FPR}^{-1}(x)\bigr)\, dx.
\end{equation*}

For all metrics, higher values indicate better detection
capability. Because vulnerability datasets are typically imbalanced (see
Table~\ref{tab:split_stats}), accuracy alone is misleading: a trivial classifier
that predicts "non-vulnerable" for every input achieves high accuracy while
detecting no vulnerabilities at all. We therefore rely primarily on F1 and AUC,
which remain informative under class imbalance, and report accuracy, precision,
and recall to fully characterize each model's behavior, including the trade-off
between false alarms and missed vulnerabilities.

We select the metrics reported for each research question according to its
specific goal:
\begin{itemize}
    \item \textbf{RQ1 (Semantic Enhancement)} concerns whether richer
    representations improve detection. We report Accuracy, F1, and AUC, and
    center the analysis on the F1 improvement ($\Delta$F1) obtained when
    replacing structural-only representations with progressively richer semantic
    embeddings: F1 captures the net effect on both false positives and false
    negatives, while AUC confirms that the improvement holds across all decision
    thresholds rather than only at the default operating point.
    \item \textbf{RQ2 (GNN Architecture)} compares architectures to identify the
    best performer on each benchmark. We use F1 as the primary selection
    criterion, since it summarizes the balance between false positives and false
    negatives in a single number at the chosen decision threshold. We
    additionally report Accuracy, AUC, and Recall: AUC to assess detection
    performance across all decision thresholds rather than at a single fixed one,
    and Recall to reveal how much of a given F1 stems from the fraction of
    vulnerabilities actually caught. On the highly imbalanced PrimeVul benchmark
    we additionally report Precision, because under such extreme imbalance
    precision is both the hardest metric to achieve and the most informative
    about a model's false-alarm burden -- the practical bottleneck for
    deployment -- and it varies substantially across architectures that
    otherwise reach comparable F1.
    \item \textbf{RQ3 (Cross-Dataset Transfer)} measures degradation under
    distribution and schema shift. We quantify it using the F1 drop, and
    equivalently the F1 retention, between in-domain and transfer settings,
    because F1 is comparable across datasets with different class distributions
    and is therefore the most meaningful basis for measuring generalization loss.
    \item \textbf{RQ4 (Multi-Dataset Generalization)} asks whether performance is
    maintained across benchmarks with diverse characteristics. We assess it
    through F1 consistency across all four datasets, as F1 remains informative
    under the widely varying class distributions, label-quality levels, and graph
    sizes that these benchmarks span.
\end{itemize}
} \section{Results}
\label{sec:results}
\changedold{This section presents our experimental results, organized by research question. The evaluation metrics and their mapping to each research question are defined in Section~\ref{sec:metrics}. A detailed comparison with prior work is provided in Section~\ref{app:baseline_comparison}.}

\subsection{RQ1: Impact of Semantic Embeddings}
\label{sec:rq1_results}

\textbf{Research Question:} \emph{To what extent do pre-trained code embeddings improve F1 score and AUC over baseline representations?}

We first establish a critical baseline: \textbf{structural-only representation}, which encodes graph topology without semantic code content. This representation used by approaches like Devign~\cite{zhou2019devign} consists of one-hot vectors indicating syntactic category (e.g., CALL, IDENTIFIER, LITERAL) from the CPG node \texttt{label} property, and categorical encoding of relationship types (e.g., AST, CFG, REACHING\_DEF) from CPG edge labels. However, no semantic content is encoded: the actual code text variable names, function names, literals extracted from the \texttt{CODE} property are discarded.

Table~\ref{tab:structural_only} shows the performance of structural-only representation across all four datasets. On the nearly-balanced Devign dataset, structural-only representation achieves \textbf{51.37 F1} barely above random chance (50\%). This demonstrates that \emph{graph topology alone, without semantic understanding of code content, is insufficient for vulnerability detection}. The model learns which node types connect via which edge types, but cannot distinguish dangerous function calls (e.g., \texttt{strcpy}) from safe alternatives (e.g., \texttt{strncpy\_s}) because both are encoded identically as CALL nodes with identical one-hot vectors. On imbalanced datasets (Draper, Reveal), accuracy appears high (87-92\%) but F1 scores (24-28\%) reveal the model's failure to identify vulnerable samples it learns to predict ``non-vulnerable'' for most inputs, exploiting class imbalance rather than learning vulnerability patterns.

\begin{table}[t]
\centering
\caption{RQ1: Structural-Only Baseline: Graph Topology Without Semantics}
\label{tab:structural_only}
\begin{tabular}{@{}lccccc@{}}
\toprule
\textbf{Dataset} & \textbf{Vuln\%} & \textbf{Acc} & \textbf{F1} & \textbf{AUC} & \textbf{vs.\ Rand.} \\
\midrule
Devign & 45.8 & 54.21 & 51.37 & 53.42 & +1.37 \\
Draper & 6.2 & 91.84 & 28.45 & 62.18 & -- \\
Reveal & 9.4 & 87.12 & 24.83 & 58.94 & -- \\
PrimeVul & 3.5 & 96.12 & 12.84 & 54.21 & -- \\
\bottomrule
\end{tabular}
\end{table}

\begin{table}[t]
\centering
\caption{RQ1: Progressive Semantic Enhancement (EdgeType-GNN, Devign)}
\label{tab:embedding_rq1}
\begin{tabular}{@{}lccccc@{}}
\toprule
\textbf{Representation} & \textbf{Dim} & \textbf{Acc} & \textbf{F1} & \textbf{AUC} & \textbf{$\Delta$F1} \\
\midrule
\multicolumn{6}{@{}l}{\textit{Structural-Only (No Semantic Content)}} \\
One-Hot Node T. & 16 & 54.21 & 51.37 & 53.42 & -- \\
\midrule
\multicolumn{6}{@{}l}{\textit{Structural + Dataset-Specific Embeddings}} \\
Word2Vec-100d & 116 & 58.34 & 62.17 & 64.89 & +10.80 \\
Word2Vec-200d & 216 & 58.92 & 63.84 & 65.72 & +12.47 \\
\midrule
\multicolumn{6}{@{}l}{\textit{Structural + Pre-trained Code Model Embeddings}} \\
CodeBERT & 784 & 59.18 & 65.21 & 67.45 & +13.84 \\
CodeT5+ & 272 & 54.62 & 60.47 & 62.83 & +9.10 \\
GraphCodeBERT & 784 & \textbf{59.52} & \textbf{66.01} & \textbf{68.34} & \textbf{+14.64} \\
UniXcoder & 784 & 59.41 & 65.73 & 67.91 & +14.36 \\
\bottomrule
\end{tabular}
\end{table}

\begin{table}[t]
\centering
\caption{RQ1: Semantic Enhancement Gain Across All Datasets}
\label{tab:semantic_gain_all}
\begin{tabular}{@{}lccccc@{}}
\toprule
 & \multicolumn{2}{c}{\textbf{Struct-Only}} & \multicolumn{2}{c}{\textbf{+GraphCodeBERT}} & \\
\cmidrule(lr){2-3} \cmidrule(lr){4-5}
\textbf{Dataset} & \textbf{Acc} & \textbf{F1} & \textbf{Acc} & \textbf{F1} & \textbf{$\Delta$F1} \\
\midrule
Devign & 54.21 & 51.37 & 59.52 & 66.01 & \textbf{+14.64} \\
Draper & 91.84 & 28.45 & 92.77 & 53.65 & \textbf{+25.20} \\
Reveal & 87.12 & 24.83 & 90.04 & 48.60 & \textbf{+23.77} \\
PrimeVul & 96.12 & 12.84 & 96.60 & 23.52 & \textbf{+10.68} \\
\bottomrule
\end{tabular}
\end{table}

\changedold{
Table~\ref{tab:embedding_rq1} reports the effect of progressively enriching the node representation on Devign, keeping the architecture fixed to EdgeType-GNN so that the contribution of the embedding is isolated from architectural effects. Each row corresponds to one representation, from the structural-only baseline through dataset-specific Word2Vec to pre-trained code models, and reports its dimensionality together with accuracy, F1, AUC, and the F1 gain over the structural-only baseline ($\Delta$F1).
The values show a consistent ordering. Dataset-specific Word2Vec already recovers a substantial part of the gap (+10.80 F1 at 100 dimensions, +12.47 at 200), indicating that even token co-occurrence statistics learned from the dataset itself carry information that one-hot type encoding discards. Pre-trained code models improve further, with GraphCodeBERT achieving the best result (66.01 F1, +14.64), closely followed by UniXcoder (+14.36) and CodeBERT (+13.84); AUC follows the same ordering, confirming that the gain is not an artifact of the decision threshold. The one exception is CodeT5+ (+9.10), which underperforms Word2Vec-200d despite being pre-trained, suggesting that its lower-dimensional (256) representation of short code fragments is less informative in this setting than a larger embedding trained directly on the target vocabulary. Overall, richer semantic representations yield monotonically better detection. \changed{This improvement is driven by the semantic information itself rather than by the larger feature size. Table~\ref{tab:embedding_rq1} allows the two factors to be separated because performance does not follow dimensionality. CodeT5+ uses 272 dimensions yet reaches only 60.47 F1, which is below Word2Vec-200d at 216 dimensions (63.84) and even below Word2Vec-100d at 116 dimensions (62.17). If additional dimensions were themselves responsible for the gain, a wider representation could not be outperformed by two narrower ones. Moving from the structural-only representation to the smallest semantic one, that is from 16 to 116 dimensions, already yields most of the improvement (+10.80 F1), whereas a further increase of almost seven times in dimensionality adds only around four points. What matters is therefore not the number of added dimensions, but the amount of discriminative information they carry about the code.}

Table~\ref{tab:semantic_gain_all} reports the results of our first research question. It compares, for each of the four datasets, the accuracy and F1-score of the model trained on structural representations alone (Struct-Only, i.e., graph structure and node types without semantic embeddings) against the same model when GraphCodeBERT semantic embeddings are added (+GraphCodeBERT) the last column reports the resulting F1 gain ($\Delta$F1). Two observations emerge from the values. First, the structural-only configuration achieves consistently low F1-scores on all four datasets (51.37 on Devign, 28.45 on Draper, 24.83 on Reveal, and 12.84 on PrimeVul), even where its accuracy appears high (e.g., 96.12 on PrimeVul). This gap between accuracy and F1 arises from class imbalance: a model can reach high accuracy by favouring the majority (non-vulnerable) class while still failing to identify the vulnerable minority, which is exactly what the low F1 reveals. Second, adding semantic embeddings improves the F1-score on every dataset, with gains of +14.64 on Devign, +25.20 on Draper, +23.77 on Reveal, and +10.68 on PrimeVul, alongside consistent (if smaller) accuracy improvements. Since the improvement appears across all four datasets, which differ in size, source, and class balance, we conclude that the limitation of structural-only representations is systematic rather than dataset-specific. Graph structure alone does not carry enough information to separate vulnerable from safe functions, and the semantic content of the code is a necessary complement to it.}

\noindent\textbf{Answer to RQ1:} Structural-only representations (one-hot node type encoding + edge type indices, without semantic content from the \texttt{CODE} property) perform at \textbf{near-random levels} on balanced datasets (51.37 F1 on Devign) and \textbf{fail catastrophically} on imbalanced datasets (24-28 F1 on Draper/Reveal). This empirically proves that \emph{graph topology alone cannot distinguish vulnerable from non-vulnerable code} the model cannot differentiate \texttt{strcpy} from \texttt{strncpy\_s} when both are encoded as identical CALL nodes.

Semantic enhancement transforms performance. Using dataset-specific Word2Vec, it provides +10-12 F1 points by capturing token co-occurrence patterns. With pre-trained code models (GraphCodeBERT), it provides +14-25 F1 points by leveraging semantic knowledge from large-scale code corpora, enabling recognition that \texttt{strcpy} is semantically related to other dangerous unbounded copy operations.

This confirms our core hypothesis: effective vulnerability detection requires \emph{both} structural understanding (CPG topology showing how code elements relate) \emph{and} semantic understanding (what those code elements mean). Structure without semantics yields near-random predictions.

\subsection{RQ2: GNN Architecture Comparison}
\label{sec:rq2_results}

\begin{table}[t]
\centering
\caption{RQ2: Architecture Comparison on Devign}
\label{tab:arch_devign}
\begin{tabular}{@{}lccccc@{}}
\toprule
\textbf{Architecture} & \textbf{EA} & \textbf{Acc} & \textbf{F1} & \textbf{AUC} & \textbf{Rec} \\
\midrule
EdgeType-GNN & \checkmark & \textbf{59.52} & \textbf{66.01} & \textbf{68.34} & 85.58 \\
Call-Aware & \checkmark & 56.63 & 65.76 & 65.84 & \textbf{90.68} \\
MultiView GGNN & \checkmark & 57.98 & 64.83 & 65.32 & 84.30 \\
GGNN (baseline) & $\times$ & 56.24 & 62.45 & 63.21 & 81.12 \\
\bottomrule
\end{tabular}
\end{table}

\begin{table}[t]
\centering
\caption{RQ2: Architecture Comparison on Draper}
\label{tab:arch_draper}
\begin{tabular}{@{}lccccc@{}}
\toprule
\textbf{Architecture} & \textbf{EA} & \textbf{Acc} & \textbf{F1} & \textbf{AUC} & \textbf{Rec} \\
\midrule
EdgeType-GNN & \checkmark & \textbf{92.77} & \textbf{53.65} & 87.28 & 64.61 \\
Call-Aware & \checkmark & 92.17 & 52.92 & \textbf{87.47} & \textbf{68.00} \\
RGAT & \checkmark & 91.32 & 49.68 & 87.36 & 66.18 \\
MultiView GGNN & \checkmark & 91.51 & 48.67 & 87.06 & 62.18 \\
GGNN (baseline) & $\times$ & 91.32 & 46.78 & 85.41 & 58.91 \\
GAT (baseline) & $\times$ & 87.40 & 37.13 & 83.59 & 57.45 \\
\bottomrule
\end{tabular}
\end{table}

\begin{table}[t]
\centering
\caption{RQ2: Architecture Comparison on Reveal}
\label{tab:arch_reveal}
\begin{tabular}{@{}lccccc@{}}
\toprule
\textbf{Architecture} & \textbf{EA} & \textbf{Acc} & \textbf{F1} & \textbf{AUC} & \textbf{Rec} \\
\midrule
MultiView GGNN & \checkmark & 90.04 & \textbf{48.60} & 84.53 & 53.06 \\
RGAT & \checkmark & \textbf{90.40} & 46.46 & 82.40 & 46.94 \\
HGT & \checkmark & 89.72 & 45.30 & \textbf{84.64} & 47.96 \\
Hypergraph & \checkmark & 88.37 & 45.89 & 83.74 & \textbf{55.61} \\
EdgeType-GNN & \checkmark & 89.45 & 44.92 & 82.33 & 48.47 \\
Call-Aware & \checkmark & 88.37 & 40.09 & 79.90 & 43.88 \\
\bottomrule
\end{tabular}
\end{table}

\begin{table}[t]
\centering
\caption{RQ2: Architecture Comparison on PrimeVul (3.5\% Vulnerable)}
\label{tab:primevul_results}
\begin{tabular}{@{}lccccc@{}}
\toprule
\textbf{Arch.} & \textbf{Acc} & \textbf{AUC} & \textbf{Prec} & \textbf{Rec} & \textbf{F1} \\
\midrule
Hypergraph & \textbf{96.60} & 81.66 & \textbf{23.78} & 23.26 & \textbf{23.52} \\
MultiView GGNN & 96.59 & 81.55 & 21.87 & 20.15 & 20.97 \\
RGAT & 94.27 & \textbf{81.89} & 15.25 & \textbf{34.07} & 21.06 \\
Improved Devign & 95.50 & 79.70 & 16.81 & 25.46 & 20.25 \\
Improved Devign & 96.06 & 81.30 & 18.34 & 21.79 & 19.92 \\
HAGNN & 95.14 & 76.43 & 16.36 & 22.53 & 18.95 \\
\bottomrule
\end{tabular}
\end{table}

\textbf{Research Question:} \emph{Among GGNN, RGAT, HGT, MultiView, and Call-Aware architectures, which achieves the highest F1 score, and do edge-type-aware models consistently outperform edge-agnostic baselines?}

Tables~\ref{tab:arch_devign}--\ref{tab:arch_reveal} present architecture comparison results on Devign, Draper and Reveal. In each Table, the EA column indicates if the model is edge-type-aware.

\noindent\textbf{Answer to RQ2:} Edge-type-aware architectures \textbf{consistently outperform edge-agnostic baselines} across  datasets. On Draper, EdgeType-GNN achieves 53.65 F1 versus 46.78 for vanilla GGNN (\textbf{+6.87 F1 points}).\\ However, the \textbf{best architecture varies by dataset}: EdgeType-GNN wins on Devign and Draper, while MultiView GGNN wins on Reveal, suggesting that architecture choice should consider dataset characteristics. MultiView's multi-perspective aggregation may better handle Reveal's higher class imbalance (9.4\% vulnerable) and larger graphs (up to 18,682 nodes).

\changedold{Table~\ref{tab:primevul_results} reports the architecture comparison on
PrimeVul~\cite{ding2024primevul}. We consider PrimeVul the most challenging
evaluation setting among our four benchmarks for three reasons. First, it is
extremely imbalanced, only 3.5\% of its functions are vulnerable, so a model
must identify a rare minority class and accuracy becomes uninformative, since a
trivial classifier that predicts ``non-vulnerable'' for every sample already
exceeds 96\% accuracy. Second, its labels were validated through multiple
rounds of expert review and the dataset was deduplicated, eliminating the label
noise and data leakage that inflate results on earlier benchmarks. Third,
precisely because of this cleaner labelling, prior work has shown that
models suffer substantial performance drops when
re-evaluated on PrimeVul~\cite{ding2024primevul}, so performance obtained here
cannot be attributed to memorising noisy or duplicated samples.

The values in Table~\ref{tab:primevul_results} reflect exactly this setting.
\textbf{Accuracy is misleading}: all models exceed 94\% accuracy, but this
mirrors the majority-class rate rather than any detection capability, which is
why we focus on AUC, precision, recall, and F1. \textbf{F1 scores are markedly
lower} (18--24\%) than on the other datasets, consistent with PrimeVul's higher
label quality and extreme imbalance rather than with a weakness specific to any
single architecture. Among the models, the \textbf{hypergraph architecture
performs best} (23.52 F1), outperforming fine-tuned code language model
baselines (18.05-21.43 F1, as reported in Table~\ref{tab:baseline_comparison}), which suggests that the higher-order relationships
captured by hyperedges help in detecting subtle vulnerability patterns.
\textbf{RGAT achieves the highest recall} (34.07\%), indicating that its
per-edge-type attention mechanism recovers more vulnerable samples, although at
the cost of precision (15.25\%).

Our graph-based models, evaluated under the same rigorous labeling, achieve better performance (up to 23.52 F1
for the hypergraph architecture). The margin is admittedly small, but it is
obtained against baselines that were optimized end-to-end for this very
benchmark, which suggests that learning from program structure yields
representations that generalize beyond memorized lexical patterns rather than
overfitting to the specific characteristics of a single dataset. PrimeVul's
rigorous labeling exposes the true difficulty of function-level vulnerability
detection, and we therefore regard the PrimeVul results as the most faithful
estimate of how our models would behave in practice.
}

\subsection{RQ3: Cross-Dataset Transfer}
\label{sec:rq3_results}

\textbf{Research Question:} \emph{What is the performance degradation when models trained on one dataset are evaluated on another, and which architectural choices minimize this degradation?}

Table~\ref{tab:cross_dataset_rq3} presents cross-dataset transfer results using the EdgeType-GNN architecture with GraphCodeBERT embeddings. 
Table~\ref{tab:transfer_by_arch} compares how different architectures handle cross-dataset transfer (training on Draper, testing on Reveal).

\begin{table}[t]
\centering
\caption{RQ3: Cross-Dataset Transfer (F1 Score)}
\label{tab:cross_dataset_rq3}
\begin{tabular}{@{}lcccc@{}}
\toprule
\textbf{Train $\rightarrow$ Test} & \textbf{Devign} & \textbf{Draper} & \textbf{Reveal} & \textbf{Avg Drop} \\
\midrule
Devign & 66.01 & 31.24 & 28.17 & --37.13 \\
Draper & 42.58 & 53.65 & 35.42 & --14.87 \\
Reveal & 38.92 & 29.87 & 48.60 & --14.21 \\
\bottomrule
\end{tabular}
\end{table}

\begin{table}[t]
\centering
\caption{RQ3: Architecture Impact on Transfer (Draper $\rightarrow$ Reveal)}
\label{tab:transfer_by_arch}
\begin{tabular}{@{}lccc@{}}
\toprule
\textbf{Architecture} & \textbf{In-Domain} & \textbf{Transfer} & \textbf{Retention} \\
\midrule
MultiView GGNN & 48.67 & 38.21 & 78.5\% \\
RGAT & 49.68 & 36.84 & 74.2\% \\
EdgeType-GNN & 53.65 & 35.42 & 66.0\% \\
Call-Aware & 52.92 & 32.17 & 60.8\% \\
\bottomrule
\end{tabular}
\end{table}

\noindent\textbf{Answer to RQ3:} Cross-dataset transfer results in \textbf{substantial performance degradation}, with an average F1 drop of 14-37 points depending on source dataset. Models trained on Devign (nearly balanced, 45.8\% vulnerable) suffer the largest transfer penalty (-37.13 avg), likely because they overfit to the balanced distribution. \\Architecturally, \textbf{MultiView GGNN exhibits the best transfer retention} (78.5\%), suggesting that multi-perspective aggregation generalizes better than single-view approaches. The schema heterogeneity (15 vs. 16 node types, 18 vs. 19 edge types) compounds with distribution shift to create a challenging transfer scenario.

\subsection{RQ4: Multi-Dataset Generalization}
\label{sec:rq4_results}

\textbf{Research Question:} \emph{Does SemVul maintain competitive performance across datasets with diverse characteristics, unlike prior approaches that excel on one benchmark but degrade on others?}

A critical limitation of existing vulnerability detection methods is their tendency to achieve strong results on a specific benchmark while performing poorly on others. This raises concerns about overfitting to dataset-specific artifacts rather than learning generalizable vulnerability patterns. We evaluate SemVul's performance consistency across four diverse benchmarks, summarized in  Table~\ref{tab:rq4_generalization}.
On a \textbf{balanced dataset (Devign, 45.8\% vuln)}, SemVul achieves 66.01 F1, a +14.64 improvement over structural-only baselines. While below the originally reported Devign score (73.26 F1), this gap is consistent with documented reproducibility issues~\cite{chakraborty2022arewethere}.
On two \textbf{moderately imbalanced datasets (Draper 6.2\%, Reveal 9.4\%)}, \changedold{SemVul achieves \textbf{state-of-the-art} results on Draper (53.65 F1, +0.45 over the RNN baseline) and competitive results on Reveal (48.60 F1, +3.33 over VulBERTa, though below the 53.09 F1 reported by Vul-LMGNNs under a possibly different split).}
Finally, on a \textbf{highly imbalanced dataset with clean labels (PrimeVul, 3.5\%)}, SemVul achieves 23.52 F1, outperforming basic fine-tuning code LM baselines (18.05-21.43 F1) by 2.09-5.47 points, demonstrating that our GNN approach competes with LLMs on rigorous benchmarks. Table~\ref{tab:baseline_comparison} reports the comparison with existing state of the art models.

\begin{table}[h]
\centering
\caption{RQ4: SemVul Performance Across Diverse Benchmarks}
\label{tab:rq4_generalization}
\begin{tabular}{@{}lcccc@{}}
\toprule
\textbf{Dataset} & \textbf{Vuln\%} & \textbf{Best F1} & \textbf{vs.\ Base} & \textbf{vs.\ Prior} \\
\midrule
Devign & 45.8 & 66.01 & +14.64 & Competitive \\
Draper & 6.2 & 53.65 & +25.20 & \textbf{+0.45} (SOTA) \\
Reveal & 9.4 & 48.60 & +23.77 & \textbf{+3.33} (vs. VulBERTa) \\
PrimeVul & 3.5 & 23.52 & +10.68 & \textbf{+2.09} (vs.\ LM) \\
\bottomrule
\end{tabular}
\end{table}

\noindent\textbf{Answer to RQ4:} Unlike prior approaches that report strong results on a single benchmark, \textbf{SemVul maintains competitive or state-of-the-art performance across all four datasets} spanning different class distributions (3.5\%-45.8\% vulnerable), label quality levels (noisy to rigorously verified), and graph sizes. This consistency demonstrates that combining CPG structural information with pre-trained semantic embeddings produces \emph{generalizable} vulnerability representations rather than dataset-specific artifacts. The semantic enhancement (RQ1) appears to be the key factor: pre-trained embeddings encode universal code understanding that transfers across benchmark characteristics.

\begin{table*}[t]
\centering
\caption{Comparison with prior work, including recent 2024--2025 approaches
that evaluate on our benchmarks ($\dagger$: results as reported in the
original papers; best variant per dataset for Vul-LMGNNs; training and test
splits may differ from ours). SemVul achieves state-of-the-art F1 on Draper,
the best F1 among directly comparable approaches on Devign, competitive
performance on Reveal, and outperforms basic fine-tuning code LM baselines on
PrimeVul. For Devign, we include the originally reported score to highlight a
known reproducibility gap in the literature.}
\label{tab:baseline_comparison}
{
\setlength{\tabcolsep}{3pt}
\renewcommand{\arraystretch}{1.1}
\begin{adjustbox}{max width=\textwidth}
\begin{tabular}{p{5.2cm} p{1.4cm} r r r r}
\hline
\textbf{Method} & \textbf{Dataset} & \textbf{Acc (\%)} & \textbf{F1 (\%)} & \textbf{ROC-AUC (\%)} & \textbf{MCC (\%)} \\
\hline
\multicolumn{6}{c}{\textbf{Devign}} \\
\hline
Devign (reported)~\cite{zhou2019devign} & Devign & 72.26 & 73.26 & - & - \\
Vul-LMGNNs$^\dagger$~\cite{liu2025vul} & Devign & 66.77 & 63.82 & - & - \\
GRACE$^\dagger$~\cite{lu2024grace} & Devign & 59.78 & 65.11 & - & - \\
\textbf{SemVul (this work)} & Devign & 59.52 & \textbf{66.01} & 68.34 & - \\
\hline
\multicolumn{6}{c}{\textbf{Draper}} \\
\hline
BOW+RF~\cite{russell2018automated} & Draper & - & 49.80 & 88.30 & 46.20 \\
RNN~\cite{russell2018automated} & Draper & - & 53.20 & 89.60 & 50.10 \\
BiLSTM~\cite{zhuang2021software} & Draper & 92.04 & 49.35 & - & 46.10 \\
CNN~\cite{zhuang2021software} & Draper & 92.26 & 49.40 & - & 46.00 \\
GGNN~\cite{zhuang2021software} & Draper & 93.49 & 50.80 & - & 47.40 \\
\textbf{SemVul (this work)} & Draper & 92.77 & \textbf{53.65} & 87.28 & - \\
\hline
\multicolumn{6}{c}{\textbf{Reveal}} \\
\hline
Baseline-BiLSTM~\cite{tian2024enhancing} & Reveal & 77.13 & 39.11 & - & - \\
Baseline-TextCNN~\cite{tian2024enhancing} & Reveal & 73.22 & 37.42 & - & - \\
REVEAL~\cite{chakraborty2022arewethere} & Reveal & 84.37 & 41.25 & - & - \\
Devign~\cite{zhou2019devign} & Reveal & 80.71 & 39.55 & - & - \\
VulBERTa~\cite{vulberta2022} & Reveal & 84.48 & 45.27 & - & - \\
GRACE$^\dagger$~\cite{lu2024grace} & Reveal & 89.73 & 43.13 & - & - \\
Vul-LMGNNs$^\dagger$~\cite{liu2025vul} & Reveal & 91.68 & \textbf{53.09} & - & - \\
\textbf{SemVul (this work)} & Reveal & 90.04 & 48.60 & 84.53 & - \\
\hline
\multicolumn{6}{c}{\textbf{PrimeVul}} \\
\hline
CodeT5 (fine-tuned)~\cite{ding2024primevul} & PrimeVul & - & 19.70 & - & - \\
CodeBERT (fine-tuned)~\cite{ding2024primevul} & PrimeVul & - & 20.86 & - & - \\
UnixCoder (fine-tuned)~\cite{ding2024primevul} & PrimeVul & - & 21.43 & - & - \\
StarCoder2 (fine-tuned)~\cite{ding2024primevul} & PrimeVul & - & 18.05 & - & - \\
CodeGen2.5 (fine-tuned)~\cite{ding2024primevul} & PrimeVul & - & 19.61 & - & - \\
\textbf{SemVul (this work)} & PrimeVul & 96.60 & \textbf{23.52} & 81.66 & - \\
\hline
\end{tabular}
\end{adjustbox}
}
\end{table*}

\subsection{Analysis of Architectural Choices}
\label{sec:additional_analysis}

We examine two key architectural choices. Regarding the \textbf{impact of GNN Depth}, Table~\ref{tab:depth_analysis} shows that shallow networks (2 layers) outperform deeper ones, aligning with the over-smoothing phenomenon where node representations converge as depth increases. As for the choice of \textbf{pooling strategy}, Table~\ref{tab:pooling_comparison} shows that multi-statistic pooling (mean + max + std) captures richer distributional information, outperforming single-statistic approaches by more than 2.5 F1 points.

\begin{table}[h]
\centering
\caption{Impact of GNN Layers (EdgeType-GNN on Devign)}
\label{tab:depth_analysis}
\begin{tabular}{lcccc}
\hline
\textbf{Layers} & \textbf{F1} & \textbf{AUC} &  \\
\hline
1 & 63.42 & 65.18  \\
2 & \textbf{66.01} & \textbf{68.34}  \\
4 & 64.87 & 66.92  \\
6 & 62.31 & 64.45  \\
\hline
\end{tabular}
\end{table}

\begin{table}[h]
\centering
\caption{Impact of Pooling Strategies (EdgeType-GNN on Devign)}
\label{tab:pooling_comparison}
\begin{tabular}{lccc}
\hline
\textbf{Pooling} & \textbf{F1} & \textbf{AUC} & \textbf{Output Dim} \\
\hline
Mean & 62.84 & 65.12 & $d$ \\
Max & 63.21 & 65.78 & $d$ \\
Mean + Max & 64.57 & 66.89 & $2d$ \\
Multi-Stat & \textbf{66.01} & \textbf{68.34} & $3d$ \\
Attention & 65.23 & 67.42 & $d$ \\
\hline
\end{tabular}
\end{table}

\subsection{Comparison with Prior Work}
\label{app:baseline_comparison}

Table~\ref{tab:baseline_comparison} compares SemVul against published baselines across all four datasets, \changedold{and more recent studies from the last two years}. Comparing our approach against the published baselines, we provide the following observations:

\begin{enumerate}
\item {Draper}: SemVul achieves 53.65 F1, outperforming the best published baseline (RNN, 53.20 F1) by 0.45 points and the graph-based GGNN baseline (50.80 F1) by 2.85 points.
\item{Reveal}: \changedold{SemVul achieves 48.60 F1, outperforming VulBERTa (45.27 F1) by 3.33 points and the original Devign model (39.55 F1) by 9.05 points}.
\item{Devign}: our 66.01 F1 falls below the originally reported 73.26 F1, consistent with reproducibility concerns documented in the literature~\cite{chakraborty2022arewethere}.
\item{PrimeVul}: SemVul achieves 23.52 F1, outperforming all basic fine-tuning code LM baselines (18.05-21.43 F1), demonstrating that GNN-based approaches can compete with large language models on high-quality benchmarks.
\end{enumerate}

\changedold{Among the more recent studies published in the last 2 years, two of them report results on the same benchmarks we use, which enables a direct positioning of our work.
GRACE~\cite{lu2024grace} evaluates on Devign and Reveal, reporting an F1 of
65.11 (accuracy 59.78) on Devign and 43.13 (accuracy 89.73) on Reveal, while
Vul-LMGNNs~\cite{liu2025vul} reports, for its best variants, an F1 of 63.82
(accuracy 66.77) on Devign and 53.09 (accuracy 91.68) on Reveal. On Devign our
best configuration achieves the highest F1 (66.01), exceeding both approaches,
and on Reveal it surpasses GRACE (48.60 vs.\ 43.13) while trailing
Vul-LMGNNs. The reported performances refers to those available in the respective  published papers, since training and test splits may differ from ours. The remaining studies are not directly
comparable. The adaptive-GNN work~\cite{liang2024source} evaluates only on
Big-Vul (F1 of 82.9), the systematic review~\cite{harzevili2024systematic}
reports no experiments of its own, Vul-LMGNNs uses a rebalanced version of
Draper (best F1 of 85.83), which is not comparable to our original imbalanced
distribution, and LLMxCPG~\cite{lekssays2025llmxcpg} evaluates PrimeVul on a
balanced paired test set (F1 of 62.06 with perfect precision) rather than the
original imbalanced protocol followed in Table~\ref{tab:primevul_results}, so
we note these protocol differences here rather than mixing incompatible
settings in one table.}

\subsection{Computational Cost}
\label{sec:computational_cost}
\changedold{
The runtime of our pipeline is best understood as a large fixed cost per function
plus a smaller cost that scales with how complex that function actually is. For
CPG generation, the fixed part comes from Joern itself because both the parse and
the export step launch a fresh Java Virtual Machine, every function pays a start-up
penalty of roughly $3.8$~s regardless of its size. Table~\ref{tab:cpg-single}
makes this concrete. Here, complexity is measured not by source length but by the
size of the resulting graph, namely its number of nodes ($V$) and edges ($E$), a
one-line function may expand into only a handful of nodes, whereas a large
control-heavy function can produce thousands.

\begin{table}[h]
\centering
\caption{Single-function CPG generation time on Devign (median of 5 repeats).}
\label{tab:cpg-single}
\begin{tabular}{lrrrrrr}
\toprule
Function & Lines & $V$ & $E$ & Parse (s) & Export (s) & Total (s) \\
\midrule
(almost empty)      & 1    & 28   & 74       & 3.36 & 1.38 & 4.74 \\
 (small)      & 5    & 62   & 220      & 2.31 & 0.97 & 3.28 \\
(largest) & 7047 & 3190 & 15{,}138 & 4.05 & 1.43 & 5.48 \\
\bottomrule
\end{tabular}
\end{table}

Notice that an essentially empty function still takes $4.74$~s, while a
$7{,}047$-line function needs only $5.48$~s, confirming that graph complexity, not
line count, is what actually drives the cost. To quantify this we swept $247$
functions across the full size range, summarised in Table~\ref{tab:cpg-scaling}, and shown graphically in Figure~\ref{fig:cpg-scaling}.
The time grows as $T \approx 3.78~\text{s} + 0.284~\text{ms}\times V$, so each
additional $1{,}000$ nodes adds only about a quarter of a second. Crucially the
growth is sub-linear (a log--log exponent of $0.66$): a $790\times$ increase in
graph size raised the time by only about $2.3\times$, and the largest graph we
encountered ($14{,}229$ nodes, $99{,}704$ edges) was still generated in $7.55$~s
rather than minutes.

\begin{table}[h]
\centering
\caption{CPG generation scaling on Devign ($n=247$ functions).}
\label{tab:cpg-scaling}
\begin{tabular}{lr}
\toprule
Quantity & Value \\
\midrule
Fixed overhead (intercept)          & about $3.78$~s \\
Marginal cost per node              & $0.284$~ms \\
Linear-fit $R^2$ (nodes / edges)    & $0.78 / 0.81$ \\
Log-log scaling exponent           & $0.66$ (sub-linear) \\
Spearman $\rho$ (nodes vs.\ time)   & $0.84$ \\
Spearman $\rho$ (lines vs.\ time)   & $0.71$ \\
Node range / time range             & $18$ to $14{,}229$ / $3.30$ to $7.62$~s \\
Largest graph ($V{=}14{,}229$, $E{=}99{,}704$) & $7.55$~s \\
\bottomrule
\end{tabular}
\end{table}

The table summarises a least-squares fit of per-function CPG generation time
against graph size, computed over a sample of $n=247$ functions that span the
full range of the Devign dataset. Each row is defined below.

\begin{description}
  \item[$n=247$ functions.] The number of functions actually measured. We
    sampled across small, medium, and large functions so the fit is not biased
    toward any one size.
  \item[$V$ (nodes) and $E$ (edges).] The two measures of graph complexity. $V$
    is the number of nodes in the generated CPG (statements, identifiers, calls,
    and so on) and $E$ is the number of edges connecting them (control-flow,
    data-flow, and syntactic relations). A larger, more complex function yields
    a larger $V$ and $E$.
  \item[$T$ (time).] The wall-clock time, in seconds, to generate one CPG,
    counting both the \texttt{joern-parse} and \texttt{joern-export} steps.
  \item[Fixed overhead (intercept), about $3.78$~s.] The constant term of the
    linear fit $T \approx \text{intercept} + \text{slope}\times V$. It is the
    time a function takes even when its graph is essentially empty, and it comes
    almost entirely from Joern starting a fresh Java Virtual Machine for each of
    its two steps. This value does not depend on function size.
  \item[Marginal cost per node, $0.284$~ms.] The slope of the same fit, that is
    the extra time added for each additional node in the graph. It is very
    small: roughly a quarter of a second for every $1{,}000$ nodes.
  \item[Linear-fit $R^2$ (nodes / edges), $0.78 / 0.81$.] The coefficient of
    determination, a goodness-of-fit score between $0$ and $1$. A value of
    $0.78$ means that graph size (node count) explains about $78\%$ of the
    variation in generation time; using edge count instead explains about
    $81\%$. Values this high indicate that graph size is a strong predictor of
    time.
  \item[Log-log scaling exponent, $0.66$ (sub-linear).] The slope obtained when
    both time and node count are plotted on logarithmic axes. It captures how
    fast time grows with size: an exponent of $1$ would mean strictly
    proportional growth, above $1$ would mean faster-than-proportional
    (super-linear), and below $1$ means slower-than-proportional (sub-linear).
    Our value of $0.66$ is comfortably below $1$, so doubling the graph size
    less than doubles the time.
  \item[Spearman $\rho$ (nodes vs.\ time), $0.84$.] Spearman's rank correlation
    coefficient, a value between $-1$ and $1$ that measures how consistently one
    quantity increases with another. A value of $0.84$ shows a strong, reliable
    tendency for generation time to rise as node count rises.
  \item[Spearman $\rho$ (lines vs.\ time), $0.71$.] The same correlation, but
    measured against source-code line count instead of node count. It is lower
    ($0.71$ versus $0.84$), which confirms that graph complexity is a better
    predictor of cost than raw source length.
  \item[Node range / time range, $18$ to $14{,}229$ / $3.30$ to $7.62$~s.] The
    smallest and largest values seen in the sample. The graphs ranged from $18$
    nodes up to $14{,}229$ nodes, yet the corresponding generation time ranged
    only from $3.30$~s to $7.62$~s, showing how weakly time responds to size.
  \item[Largest graph ($V{=}14{,}229$, $E{=}99{,}704$), $7.55$~s.] The single
    biggest CPG in the sample, reported separately to make the point explicit:
    even a graph with over $14{,}000$ nodes and nearly $100{,}000$ edges was
    generated in about seven and a half seconds.
\end{description}

\begin{figure}
\centering
\begin{minipage}[t]{0.48\textwidth}
  \centering
  \includegraphics[width=\linewidth]{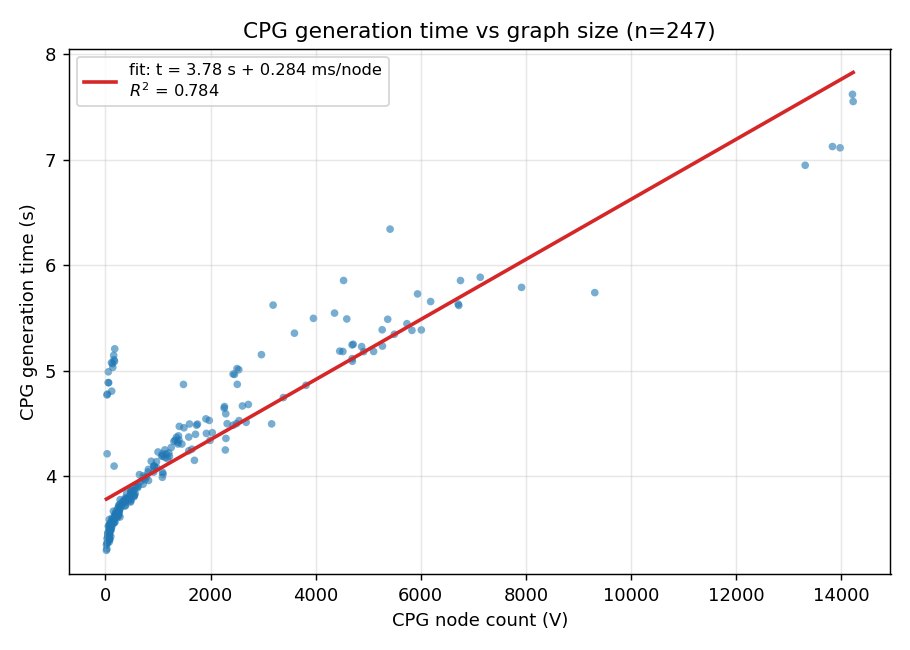}
\end{minipage}\hfill
\begin{minipage}[t]{0.48\textwidth}
  \centering
  \includegraphics[width=\linewidth]{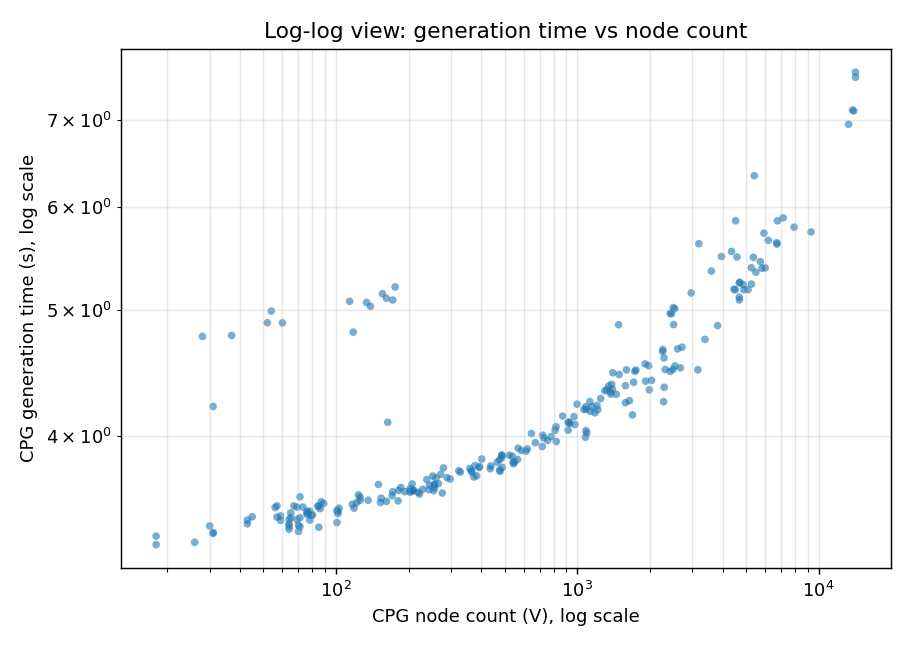}
\end{minipage}
\caption{CPG generation time versus graph size on Devign ($n=247$ functions).
Left: linear scale with the fitted line
$T \approx 3.78\,\text{s} + 0.284\,\text{ms}\times\lvert V\rvert$
($R^2=0.78$); the intercept is the fixed Joern/JVM overhead and the cluster of
small functions sits on that floor. Right: the same data on logarithmic axes;
the flat region at small node counts is the overhead-dominated regime, and the
gentle slope beyond it reflects the sub-linear (exponent ${\approx}0.66$)
growth of the complexity term.}
\label{fig:cpg-scaling}
\end{figure}

Finally, we measured how long the trained model takes to classify a single
unseen function, walking up the complexity scale from the smallest function in
our measured sample to the largest. All three are real Devign functions; graphs
were built at exactly their measured node/edge counts and passed through the
trained architecture one at a time (batch size~1) on GPU (Table~\ref{tab:inference}).\\

\begin{table}[t]
\centering
\caption{Per-function testing time on Devign. End-to-end is CPG generation plus inference.}
\label{tab:inference}
\begin{tabular}{lrrrrrr}
\toprule
Function & Lines & $V$ & $E$ & Inference & CPG gen. & End-to-end \\
\midrule
Smallest & 1     & 18       & 42       & 1.60\,ms  & 3.30\,s & 3.30\,s \\
Medium   & 70    & 487      & 3{,}018  & 1.95\,ms  & 3.87\,s & 3.87\,s \\
Largest  & 3390  & 14{,}229 & 99{,}704 & 16.48\,ms & 7.55\,s & 7.57\,s \\
\bottomrule
\end{tabular}
\end{table}

Testing behaves the same way once the model is trained. Walking up the
complexity scale, the smallest function in our sample, a one-line arithmetic
operation whose graph has just 18 nodes, is classified in $1.60$~ms; the medium
function, a realistic 70-line function sitting at the dataset median (487 nodes,
about $3{,}000$ edges), needs only $1.95$~ms despite having $70\times$ the lines
and $72\times$ the edges, because at this scale the GPU's fixed kernel-launch
overhead of roughly $1.6$~ms dominates, much like the JVM start-up does for
Joern; and even the largest graph we encountered ($14{,}229$ nodes, $99{,}704$
edges, about $33\times$ the edges of the medium one) takes just $16.48$~ms, a
factor of $8.5$ rather than $33$, so the growth is once again sub-linear. In
other words, testing time depends somewhat on how long a function is, but far
more directly on how complex its graph turns out to be, and it is never the
bottleneck. Even in the worst case inference stays below $17$~ms, around
$0.2\%$ of the end-to-end testing time, with the remainder being the one-off
CPG generation for that new function ($3.3$ to $7.6$~s).\\
The other two stages
are similarly well-behaved. Embedding is precomputed once, and the related cost
depends on the number of unique code strings rather than the dataset size,
since identical strings are embedded once and cached; at the roughly $4{,}600$
embeddings per second we measured on GPU, even hundreds of thousands of unique
strings take a couple of minutes and are then reused across every training run.
Training itself is governed by the number of graphs, their average size, and
the model configuration (hidden dimension, layers, and edge types), coming to
about $29$~s per epoch and roughly $39$ minutes for a full run on Devign.
Hardware naturally plays a role throughout: the CPG stage is CPU and JVM bound
and is therefore the slowest per item, while embedding, training, and inference
all run on the GPU, where they are substantially faster than on CPU. Taken
together, none of the stages scales badly with function complexity, a
brand-new, never-seen function goes from raw source to vulnerability verdict in
under $8$~s even in the worst case, and the dominant cost factor throughout is
simply the number of functions in the dataset, which is exactly the behaviour
one wants from a model intended for real-world use such as continuous
integration or code review.

\begin{table}
\centering
\caption{GNN training and testing cost on Devign}
\label{tab:train}
\begin{tabular}{lr}
\toprule
Metric & Value \\
\midrule
Per epoch ($21{,}854$ graphs) & about $29$~s (train $\sim$27.4, val $\sim$1.6) \\
Throughput                    & about $750$ graphs/s \\
Full run ($\sim$80 epochs)    & about $39$ minutes \\
\midrule
Test, smallest function ($V{=}18$, $E{=}42$)        & $1.60$~ms \\
Test, medium function ($V{=}487$, $E{=}3{,}018$)    & $1.95$~ms \\
Test, largest function ($V{=}14{,}229$, $E{=}99{,}704$) & $16.48$~ms \\
End-to-end test of a new function      & $3.3$ to $7.6$~s \\
\bottomrule
\end{tabular}
\end{table}
}

\subsection{Summary of Findings}
\label{sec:results_summary}

Table~\ref{tab:rq_summary} summarizes our answers to the four research questions. These findings suggest prioritizing embeddings over architecture complexity, as pre-trained code models provide larger gains than architectural innovations. Shallow networks (2 layers) outperform deeper models while reducing training cost. Architecture should be matched to dataset characteristics: MultiView GGNN handles class imbalance better, while EdgeType-GNN excels on balanced data. Finally, transfer penalties should be expected; practitioners should deploy models on data similar to the training distribution or employ domain adaptation techniques.

\begin{table}[t]
\centering
\caption{Summary of Research Question Answers}
\label{tab:rq_summary}
\begin{tabular}{p{1.2cm}p{11cm}}
\hline
\textbf{RQ} & \textbf{Key Finding} \\
\hline
RQ1 & Structural-only representation (one-hot node types + edge types) achieves \textbf{near-random F1} (51.37\% on Devign, 12.84\% on PrimeVul). Pre-trained embeddings improve F1 by \textbf{+10-25 points} across all datasets. \\
\hline
RQ2 & Edge-type-aware architectures consistently outperform edge-agnostic baselines (\textbf{+6.87 F1} on Draper). The optimal architecture varies by dataset: EdgeType-GNN for balanced datasets, Hypergraph/MultiView for imbalanced ones. \\
\hline
RQ3 & Cross-dataset transfer incurs \textbf{14-37 F1 point degradation}. MultiView GGNN achieves best transfer retention (78.5\%). Schema heterogeneity and distribution shift compound to create challenging transfer scenarios. \\
\hline
RQ4 & Unlike prior methods that excel on one benchmark, SemVul achieves \textbf{competitive or SOTA results across all four datasets} (Devign, Draper, Reveal, PrimeVul) with diverse characteristics (3.5\%-45.8\% vulnerable), demonstrating generalizable vulnerability detection. \\
\hline
\end{tabular}
\end{table} \section{Conclusion}
\label{sec:conclusion}

We presented SemVul, a vulnerability detection approach that combines Code Property Graphs with pre-trained code embeddings. Our evaluation across four datasets and six GNN architectures reveals that graph topology alone yields near-random performance, while semantic embeddings provide +10-25 F1 points improvement. Edge-type-aware architectures consistently outperform edge-agnostic baselines, and cross-dataset transfer incurs 14-37 F1 point penalties, with MultiView architectures exhibiting better retention. SemVul achieves state-of-the-art results on Draper (53.65 F1), \changedold{competitive results on Reveal (48.60 F1)}, and outperforms code LM baselines on PrimeVul (23.52 vs.\ 18.05-21.43 F1). 
This consistent performance across diverse benchmarks from balanced Devign (66.01 F1) to highly imbalanced PrimeVul validates our core hypothesis: effective vulnerability detection requires both structural understanding from CPG topology and semantic understanding from pre-trained embeddings. Our reproducible methodology provides a foundation for future work on cross-language transfer, interprocedural analysis, and real-world deployment.

\changedold{
\subsection{Interpretability and Practical Use}
In its present form, SemVul is intended as a detection aid rather than a fully autonomous decision-maker. The model flags candidate vulnerable functions, narrowing the portion of a codebase that a security analyst must examine, while the final judgment on whether a flagged fragment constitutes an exploitable vulnerability remains with the human expert. This positioning is deliberate: given the difficulty of the task on rigorously labeled benchmarks (e.g., the low F1 scores observed on PrimeVul), automated predictions are best treated as prioritization signals that focus expert effort rather than as conclusive verdicts.
A promising direction for increasing the practical value of the approach is explainability. Unlike token-based or LLM-based detectors, whose reasoning is difficult to localize in the source code, the CPG-based representation used by SemVul is inherently structural: a prediction can, in principle, be attributed to specific nodes and edges of the graph, such as the data-flow chain linking an untrusted input to a dangerous library call, or a control-flow region lacking a protective conditional. In future work, we plan to incorporate explanation mechanisms into the pipeline — for instance, by exploiting the attention weights of edge-type-aware architectures or by extracting the subgraphs most responsible for a given prediction — so that SemVul can not only detect a vulnerability but also indicate the code elements that motivated its decision.} 
\section*{Declaration of generative AI and AI-assisted technologies in the manuscript preparation process}
During the preparation of this work the authors used OpenAI ChatGPT and Superhuman Grammarly to support grammar and language checking, and to assist with rephrasing and improving the clarity and readability of the manuscript. After using these tools, the authors reviewed and edited the content as needed and take full responsibility for the content of the published article.

\section*{Artifact Availability}
All code, scripts, and related documentation necessary to reproduce our pipeline are available at:
\begin{center}
\url{https://github.com/Affan47/SemVul}
\end{center}
In particular, the \texttt{README.md} file in the main folder of the repository illustrates the steps necessary to reproduce the experimental results detailed in this paper.

\section*{Funding}
This work was partially supported by project SERICS (PE00000014) under the Italian NRRP MUR programme funded by the European Union - NextGenerationEU.

\bibliographystyle{cas-model2-names}
\bibliography{refs}

\appendix
\section{CPG Schema Details}
\label{app:schemas}

\begin{table*}[t]
\centering
\caption{CPG Node Types and Their Vulnerability Relevance}
\label{tab:node_types}
\begin{adjustbox}{max width=\textwidth}
\begin{tabular}{llcp{8cm}}
\hline
\textbf{Node Type} & \textbf{Category} & \textbf{Datasets} & \textbf{Description and Vulnerability Relevance} \\
\hline
CALL & Invocation & All & Function calls; most vulnerability-critical (dangerous APIs like \texttt{strcpy}, \texttt{sprintf}) \\
IDENTIFIER & Reference & All & Variable references; track data flow from tainted sources \\
LITERAL & Constant & All & Constant values including buffer sizes \\
BLOCK & Structure & All & Code block containers \\
METHOD & Entry & All & Function entry points defining analysis scope \\
LOCAL & Declaration & All & Local variable declarations including buffer allocations \\
CONTROL\_STRUCTURE & Control & All & Conditionals and loops; presence determines safety checks \\
RETURN & Exit & All & Function exit points; relevant for resource leak detection \\
FIELD\_IDENTIFIER & Reference & All & Struct/class field references \\
TYPE\_DECL & Declaration & All & Type declarations \\
JUMP\_TARGET & Control & All & Labels and jump destinations \\
METHOD\_REF & Reference & All & References to methods/functions \\
UNKNOWN & Other & All & Unclassified nodes \\
DEPENDENCY & Import & All & External dependencies \\
IMPORT & Import & All & Import statements \\
MEMBER & Reference & Reveal, Draper & Struct/class member access patterns \\
\hline
\end{tabular}
\end{adjustbox}
\end{table*}

\begin{table*}[t]
\centering
\caption{CPG Edge Types and Their Vulnerability Relevance}
\label{tab:edge_types}
\begin{adjustbox}{max width=\textwidth}
\begin{tabular}{llcp{8cm}}
\hline
\textbf{Edge Type} & \textbf{Source Graph} & \textbf{Datasets} & \textbf{Description and Vulnerability Relevance} \\
\hline
\multicolumn{4}{l}{\textit{Data-Flow Edges ($G_{\text{DFG}}$)}} \\
REACHING\_DEF & DFG & All & Variable definition reaches use; critical for taint tracking \\
\hline
\multicolumn{4}{l}{\textit{Control-Flow Edges ($G_{\text{CFG}}$)}} \\
CFG & CFG & All & Execution order; essential for detecting missing checks \\
DOMINATE & CFG & All & Node dominates another in control flow \\
POST\_DOMINATE & CFG & All & Node post-dominates another \\
\hline
\multicolumn{4}{l}{\textit{Syntax Edges ($G_{\text{AST}}$)}} \\
AST & AST & All & Parent-child syntax relationships \\
\hline
\multicolumn{4}{l}{\textit{Program Dependence Edges ($G_{\text{PDG}}$)}} \\
CDG & PDG & All & Control dependence; conditional execution \\
\hline
\multicolumn{4}{l}{\textit{Call and Reference Edges}} \\
CALL & Call Graph & All & Function invocation relationship \\
ARGUMENT & Call Graph & All & Argument passed to function \\
REF & Reference & All & Reference relationship \\
PARAMETER\_LINK & Call Graph & All & Parameter binding \\
RECEIVER & Call Graph & All & Method receiver object \\
\hline
\multicolumn{4}{l}{\textit{Type and Binding Edges}} \\
EVAL\_TYPE & Type & All & Expression type evaluation \\
BINDS & Binding & All & Variable binding \\
ALIAS\_OF & Alias & All & Aliasing relationship \\
\hline
\multicolumn{4}{l}{\textit{Structural Edges}} \\
CONDITION & Control & All & Condition expression \\
CONTAINS & Structure & All & Containment relationship \\
SOURCE\_FILE & Structure & All & Source file association \\
IMPORTS & Import & All & Import relationship \\
CAPTURE & Closure & Reveal, Draper & Variable capture in closures \\
\hline
\end{tabular}
\end{adjustbox}
\end{table*}

This appendix provides the complete node and edge type vocabularies discovered from each dataset's CPG exports. In particular, Table~\ref{tab:node_types} describes the CPG node types and their roles in vulnerability detection, while Table~\ref{tab:edge_types} describes the CPG edge types organized by their source graph representation.

\end{document}